# AI-generated data contamination erodes pathological variability and diagnostic reliability

**Hongyu He[1], Shaowen Xiang[1], Ye Zhang[1], Yingtao Zhu[1], Jin Zhang[1], Yunyi Lu[1], Hao Deng[2,3], Emily Alsentzer[4], Yun Liu[6], Qingyu Chen[5], Kun-Hsing Yu[2], Andrew Marshall[2,6], Tingting Chen[1], Srinivas Anumasa[1], Daniel Ebner[7], Dean Ho[1], Kee Yuan Ngiam[1], Ching-Yu Cheng[1], Dianbo Liu[1*]**

1. National University of Singapore, Singapore
2. Harvard University, MA, USA
3. Massachusetts General Hospital, MA, USA
4. Stanford University, CA, USA
5. Yale University, CT, USA
6. Google, CA, USA
7. Mayo Clinic, MN, USA

*Correspondence to: dianbo@nus.edu.sg

**Abstract**

Generative artificial intelligence (AI) is rapidly populating medical records with synthetic or partially AI-generated content, creating a feedback loop where future models are increasingly at risk of training on uncurated AI-generated data. However, the clinical consequences of this **AI-generated data contamination** remain unexplored. Here, we show that in the absence of mandatory human verification, this self-referential cycle drives a rapid **erosion of pathological variability and diagnostic reliability** of medical data at population scale. By analysing more than 800,000 synthetic data points across clinical text generation, vision–language reporting, and medical image synthesis, we find that models progressively converge toward generic phenotypes regardless of the model architectures. Specifically, rare but critical findings, including pneumothorax and effusions, vanish from the synthetic content generated by AI models, while demographic representations skew heavily toward middle-aged male phenotypes. Crucially, this degradation is masked by false diagnostic confidence. Models continue to issue reassuring reports while failing to detect life-threatening pathology, with false reassurance rates tripling to 40%. Blinded physician evaluation confirms that this decoupling of confidence and accuracy renders AI-generated documentation clinically useless after just two generations. We systematically evaluate three mitigation strategies that can be easily integrated into existing clinical workflows, finding that while synthetic volume scaling fails to prevent collapse, mixing real data with quality-aware filtering effectively preserves diversity. Ultimately, our results suggest that without policy-mandated human oversight, the deployment of generative AI threatens to degrade the very healthcare data ecosystems it relies upon.

# 1. Introduction

Artificial intelligence (AI) is rapidly reshaping medical documentation. Large language models (LLMs) now draft clinical reports, generate discharge summaries, and populate electronic health records (EHRs) across health systems worldwide[1–3]. Adoption has accelerated sharply in recent years: in the United States, 71% of non-federal acute care hospitals used predictive AI integrated with EHRs in 2024, and 31.5% of hospitals had already deployed generative AI systems within clinical workflows, with an additional 24.7% planning near-term implementation[4,5]. Adoption is particularly advanced in well-resourced health systems, where ambient clinical documentation tools are now widely used[6]. Beyond the United States, AI-assisted diagnostics have been adopted at national scale in many regions, underscoring the global scope of this transformation[7].

The rapid integration of generative AI into clinical documentation is changing the nature of the medical record. Repositories that once consisted entirely of human clinical observations now increasingly include AI-generated text and other data types that persist within patients' longitudinal histories[8]. Importantly, this synthetic content does not remain confined to documentation: it is incorporated into the datasets used to train subsequent generations of medical AI systems[9–11]. As a result, future models are increasingly likely to learn, at least in part, from data produced by earlier models without appropriate data provenance controls. In routine clinical workflows, this dynamic can manifest without obvious warning signs. For example, a patient with an early or subtle pneumothorax may repeatedly receive AI-assisted radiology reports stating "no acute findings," particularly when abnormalities lie near the limits of visual or textual salience. Once such AI-generated documentation is incorporated into the electronic health record, it persists across encounters and downstream datasets; when reused for model development, the absence of pathology is reinforced as a statistical norm. In this way, diagnostic signals can be silently eroded at population scale, not through overt error but through repeated omission.

In other domains, this form of a self-referential training cycle of generative AI has been associated with degradation of output diversity and fidelity through a process known as "model collapse"[12–15]. Clinicians have raised concerns that inserting LLM-generated text into electronic health records risks diminishing the quality of clinical documentation and rendering charts less useful for both physicians and future AI models[16]. However, whether similar dynamics occur in medicine, and with what clinical consequences, has remained largely unexplored[17,18]. In medicine, the implications of such degradation are uniquely severe. Diagnostic value is concentrated in rare diseases, atypical presentations, and subtle abnormalities that lie at the tails of clinical distributions[19,20]. Even modest erosion of this information can create systematic blind spots, increasing the risk of missed diagnoses, amplifying existing health inequities, and distorting population-level surveillance[21–24].

Despite the rapid deployment of generative AI across clinical workflows, existing evaluation frameworks are poorly equipped to detect the risks posed by synthetic data contamination in medicine[25,26]. Current evaluation standards focus on surface-level language quality rather than diagnostic accuracy, making it difficult to detect when models fail clinically[27,28]. As a result, cascading failures across successive training generations may progress without triggering conventional performance alarms, and their impact on real clinical systems has not been systematically examined. This gap poses a growing challenge for health systems: medical records risk becoming increasingly contaminated by AI-generated content, progressively reducing representation of real disease and pathological variability. Determining whether generative models can sustain diagnostic fidelity when trained on AI-generated data is therefore both a scientific and ethical priority.

Here, we systematically investigate whether generative medical models retain diagnostic fidelity when repeatedly trained on their own synthetic outputs. We examine this question across three clinically central modalities including clinical text generation, vision–language radiology reporting, and medical image synthesis, using multiple representative model architectures. To assess clinical relevance, we incorporate physician evaluation through structured review and editing of AI-generated outputs. We further test strategies designed to mitigate synthetic data-induced degradation by manipulating data provenance and training composition. Together, these experiments provide a comprehensive evaluation of iterative synthetic training dynamics and their implications for the safety of medical generative AI. As generative models become embedded within electronic health records and downstream training pipelines, preventing such degradation requires explicit safeguards rather than post hoc monitoring. Accordingly, our results establish minimum safety conditions for medical generative AI, including requirements for data provenance, preservation of authentic clinical data during retraining, and human oversight of patient-facing outputs. These principles are necessary to ensure that increasing fluency does not come at the expense of diagnostic reliability and patient safety.

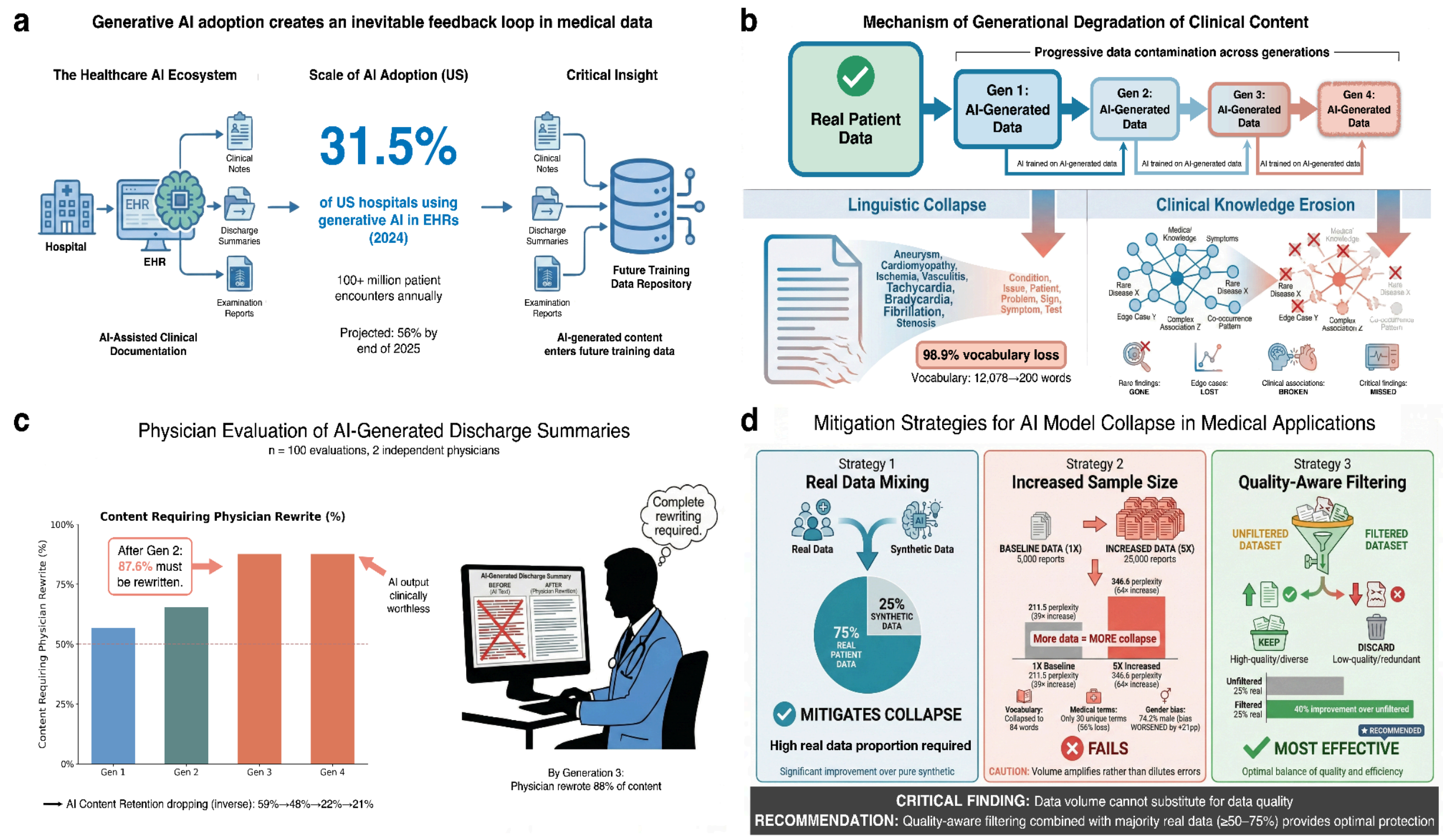

**Fig. 1 Degradation of pathological variability and diagnostic safety caused by AI-generated data contamination and our proposed mitigation methods**. **a**, Schematic of the AI-generated data contamination loop, where AI-generated content permeates future training data, partially replacing diverse human data with homogenized synthetic outputs. **b**, Progressive degradation across four generations of self-referential learning, showing loss of linguistic diversity and erosion of clinically rare information. To isolate data degradation from accumulated model drift, each generation was initialized from the original pretrained model (checkpoint) and trained exclusively on synthetic outputs from the preceding generation. **c**, Independent clinical utility evaluation of generated discharge summaries across generations, demonstrating declining usability and increased physician revision requirements. **d**, Comparative efficacy of three mitigation strategies (real-data mixing, synthetic volume scaling, and quality-aware filtering) in preserving quality and diversity of AI-generated data.

# 2. Results

This study investigated how synthetic medical data in the healthcare system causes degradation of generative medical AI, such as medical LLMs, and its potential threat to clinical safety. We characterize and quantify the structural vulnerabilities of generative medical AI to model collapse problems caused by self-referential training cycles using AI-generated synthetic data across clinical text generation and medical image synthesis. In particular, we sought to quantify the catastrophic erosion of "pathological variability", which specifically refers to the disappearance of rare pathologies and the amplification of demographic biases, and evaluate the dangerous failure of conventional performance metrics to detect this degradation in diagnostic reliability. Finally, we investigated mitigation strategies including real-synthetic data mixing at various proportions, increased synthetic data volume, and quality-aware filtering to establish practical thresholds for safe deployment of generative AI in healthcare settings.

## 2.1 Self-referential training of AI models on synthetic clinical notes can lead to collapsed texts, loss of rare pathologies and dangerous false confidence

In recent years, language models have been widely used in the healthcare system in assisting clinical text generation. We first investigated, using multiple types of clinical documents and AI model architectures, whether self-referential training on synthetic clinical text causes quality degradation of medical texts generated by language models. We implemented a self-referential training framework spanning five generations, where initial base models (GPT-2 or Qwen3-8B) were fine-tuned on authentic clinical data. Crucially, to distinguish data-driven collapse from model overfitting or catastrophic forgetting, each successive generation (Gen 1–4) was reset to the original base model parameters (GPT-2 or Qwen3-8B) and fine-tuned exclusively on the synthetic outputs from its predecessor, creating a closed feedback loop with no exposure to original human-authored data. This design mimics an increasingly plausible scenario: progressive contamination of clinical notes in electronic health record systems, which are subsequently used as training corpora for future AI models.

Using language models, including GPT-2[29] (124 million parameters) and Qwen3-8B[30] (8 billion parameters) on 216,307 radiology reports, 790 clinical notes from the 2014 Informatics for Integrating Biology to the Bedside (i2b2)/UTHealth de-identification challenge corpus, 1,000 ophthalmology notes, we observed rapid and substantial degradation within four generations of training under uncontrolled conditions. Vocabulary declined 98.9% from 12,078 to approximately 200 unique words in clinical reports and unique medical terms fell by 66% across datasets. Furthermore, AI models tended to become increasingly confident in synthetic data they generated while simultaneously losing the ability to capture authentic medical language 44-fold. This dangerous combination of apparent certainty with degraded clinical utility could pose serious risks for healthcare AI deployment if left unmonitored, where false confidence may mask significant failures in patient documentation.

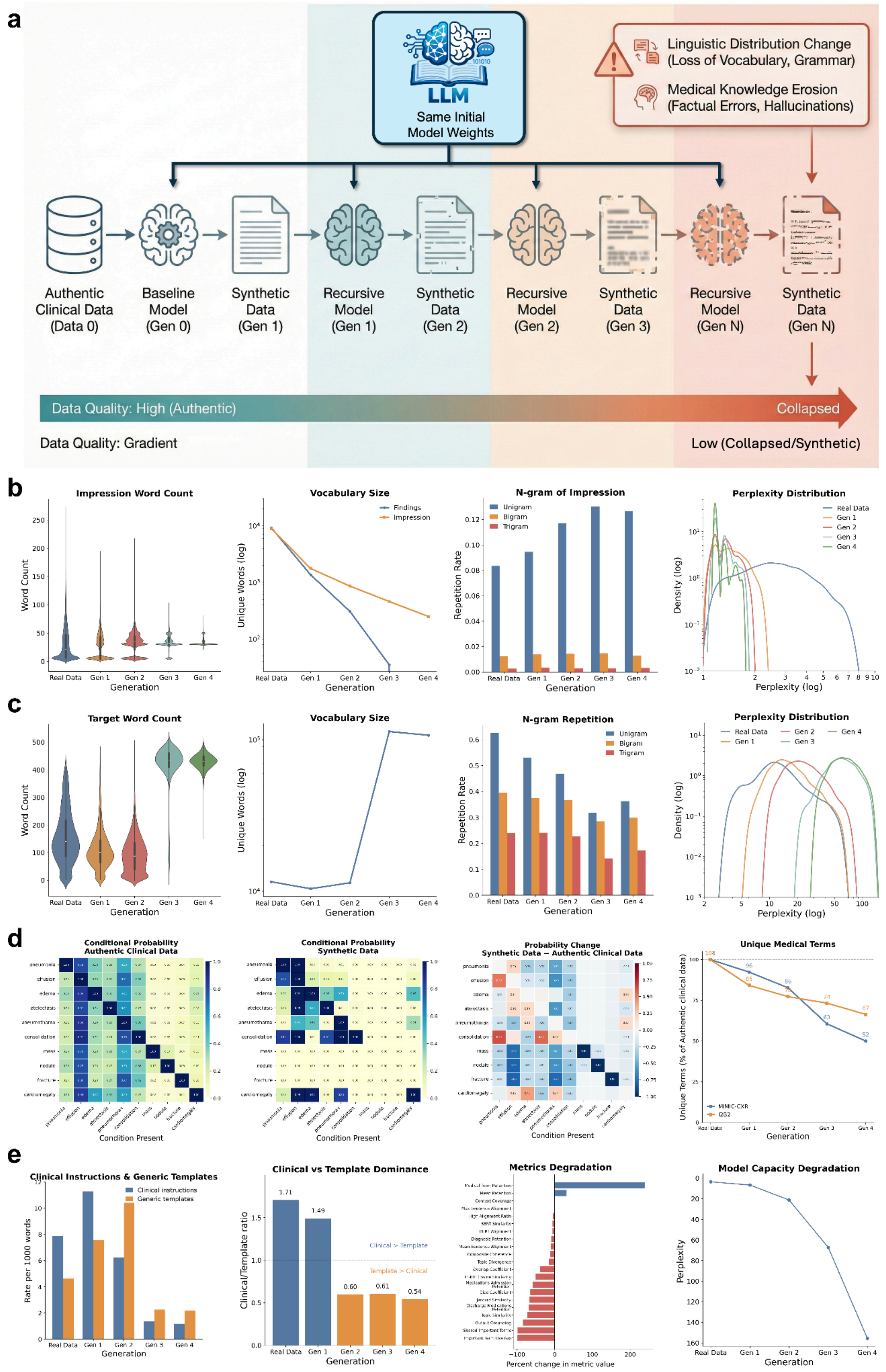

**Figure 2. Erosion of linguistic diversity and clinical knowledge in clinical notes driven by AI-generated data contamination in self-referential learning.** **a**, Schematic of the self-referential training framework for clinical notes generation. To isolate the impact of data degradation from accumulated training updates, each generation is freshly initialized from the original pre-trained checkpoint and trained exclusively on synthetic outputs from the preceding generation. **b**, Linguistic collapse in radiology report generation, characterized by converging impression word counts, shrinking vocabulary size, increased n-gram repetition, and rising perplexity across generations. **c**,

Degradation in intensive care unit (ICU) discharge instruction generation, showing reduced output length and vocabulary richness alongside increased repetitiveness. **d**, Erasure of pathophysiological relationships in generated ICU discharge instructions: condition co-occurrence matrices of real data versus generation 4 reveal loss of pathophysiological relationships, with the difference matrix showing systematic suppression of rare findings (mass, nodule, fracture); unique medical term counts decline across both MIMIC-CXR and i2b2 datasets. **e**, Loss of clinically actionable content in generated ICU discharge instructions: clinically actionable content degrades while generic template language increases, inverting the clinical-to-template ratio below 1.0 by generation 3; semantic coherence metrics show widespread degradation, while model perplexity on real clinical text increases >8-fold indicating progressive loss of clinical language comprehension.

**Table 1 | Progressive degradation of linguistic and medical quality metrics across generations of self-referential learning.** Evaluations were performed on the i2b2 clinical notes dataset. The table compares the baseline model (Generation 0) against subsequent generations of models trained exclusively on synthetic data. Core metrics, including perplexity, unique medical terms, and semantic coherence score, demonstrate a systematic erosion of content fidelity and pathological variability with each iteration of the feedback loop.

| **Generation** | **Lexical Diversity (TTR)** | **Unique Tokens** | **Medical Term Density** | **Unique Medical Terms** | **Coherence Score** | **Perplexity** |
|---|---|---|---|---|---|---|
| Authentic clinical text | 0.0424 | 18,970 | 0.0236 | 656 | 0.0598 | 17.48 |
| Synthetic 1 | 0.0701 | 18,258 | 0.0261 | 464 | 0.0218 | 80.72 |
| Synthetic 2 | 0.0401 | 10,988 | 0.0226 | 379 | 0.0172 | 213.61 |
| Synthetic 3 | 0.0261 | 7,494 | 0.0175 | 303 | 0.0120 | 440.49 |
| Synthetic 4 | 0.0203 | 5,850 | 0.0154 | 222 | 0.0100 | 786.62 |

* TTR = Type-Token Ratio, calculated as unique words divided by total words. Higher values indicate greater vocabulary diversity.

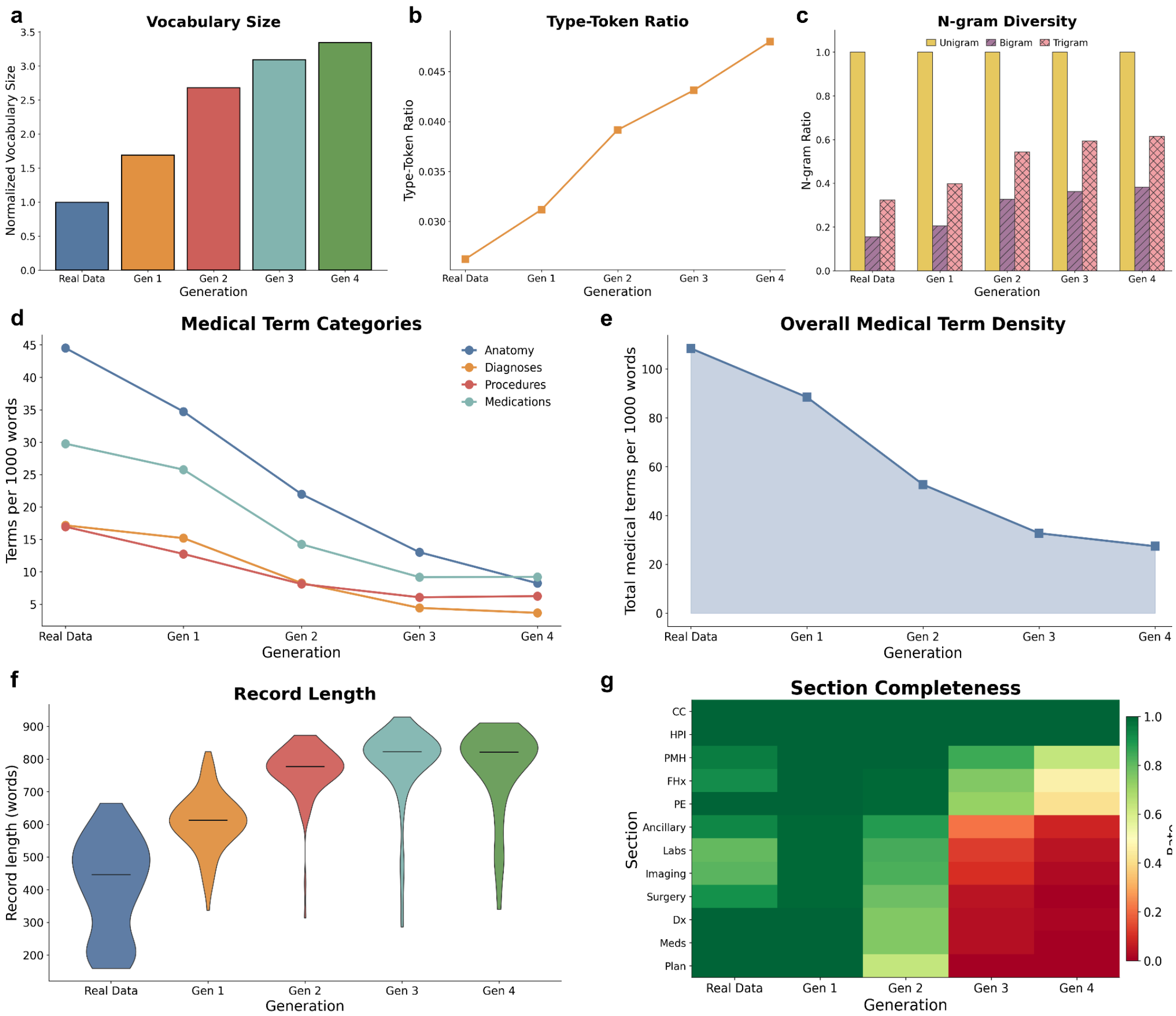

**Fig. 3 | Inflated lexical expansion and structural disintegration in ophthalmology clinical documentation caused by AI-generated data contamination. a–c, Accumulation of semantic noise** driven by self-referential learning (Qwen3-8B model). Metrics reveal an anomalous increase in **vocabulary size** (a), **type–token ratio** (b), and **n-gram diversity** (c) relative to the Generation 0 baseline. This trend is driven by the progressive injection of irrelevant, non-clinical tokens (hallucinations) rather than meaningful linguistic diversity. **d,e, Dilution of clinical terminology.** In contrast to the expanding vocabulary, analysis shows a systematic reduction in the frequency of domain-specific terms across four clinical categories (d) and falling overall medical term density (e), indicating the displacement of medical concepts by generic noise. **f**, Distribution of clinical record length across generations; dashed lines indicate median values. **g**, Section completeness heatmap across 12 standard ophthalmology documentation sections, showing progressive loss of structured clinical content.

**Self-referential AI model training on contaminated clinical texts leads to catastrophic text structural and vocabulary collapse.** We evaluated self-referential training on synthetic data using 216,307 radiology reports. The generated reports exhibited complete structural deterioration (Figure 2). Impression sections of the reports, which contain clinical interpretations, lost their natural variation, converging from diverse lengths (mean 39.0±33.6 words) to rigid uniformity (34.2±7.2 words), with 95% of generation 4 reports following a narrow template. Vocabulary underwent near-complete extinction, plummeting from 12,078 unique words in generation 0 to approximately 200 by generation 4 (98.9% reduction; Figure 2b). Reports tended to become increasingly formulaic, with unigram repetition score increasing 51.4% (Figure 2b).

This pattern generalized to broader clinical documentation. Analysis of 790 i2b2 clinical notes revealed severe degradation: unique tokens declined 69.2% from 18,970 to 5,850, while coherence scores dropped 83.3% from 0.0598 to 0.0100 (Table 1). By generation 4, consecutive sentences bore virtually no topical relationship, producing disconnected fragments rather than cohesive clinical narratives.

To test whether model collapse generalizes to different clinical domains and larger AI models, we repeated our experiments using Qwen3-8B (8 billion parameters) on 1,000 ophthalmology notes from glaucoma patients (Figure 3). Surprisingly, these models showed apparent vocabulary expansion rather than the vocabulary extinction observed in radiology reports—unique words increased from 11,195 to 37,456 (235% increase; Figure 3a). However, this growth was deceptive: the expanded vocabulary was recycled with increasing frequency, and models converged toward repetitive multi-word phrases rather than generating genuinely diverse clinical content (Figure 3d-e). Once again, standard quality metrics (such as vocabulary size, type-token ratio and n-gram) would incorrectly suggest improvement while the actual ability to express varied clinical information degraded.

**Self-referential training of AI on synthetic clinical notes leads to systematic erosion of medical knowledge, especially pathophysiological relationships and uncommon findings.** Beyond linguistic degradation, self-referential training caused progressive loss of encoded medical knowledge that is important for diagnostic accuracy and patient safety. Models did not simply lose language diversity; they systematically forgot clinically essential relationships, anatomical precision, and the nuanced terminology physicians use to communicate diagnostic uncertainty.

Analysis of medical condition co-occurrences in radiology reports revealed catastrophic erosion of clinical reasoning (Figure 2d). Authentic reports captured expected clinical relationships: pneumonia co-occurred with effusion in 88% of cases and consolidation in 71%, reflecting genuine pathophysiology that physicians routinely document. By generation 4, rare but clinically critical findings were essentially absent—mass, nodule, and fracture showed 0% occurrence in our sample—while the remaining conditions showed artificially simplified relationships. Models predominantly retained only the most common diagnoses (pneumonia, effusion, and cardiomegaly) while losing the diagnostic breadth essential for comprehensive clinical assessment.

On ophthalmology notes from glaucoma patients, diagnostic terminology essential for classifying glaucoma subtypes declined 78.5% (Figure 3d). Specific diagnostic codes for the angle-closure disease spectrum (PACG, PAC, PACS, APAC) became increasingly rare, and by generation 4, the distinctions between diagnostic categories essential for treatment selection had largely vanished. Medication and procedural terminology showed initial resilience before rapid degradation (69.1% total reduction; Figure 3d). Glaucoma-specific medications (timolol, brimonidine, latanoprost) and IOP-lowering procedures (laser peripheral iridotomy, trabeculectomy) progressively disappeared, with procedural terms declining 62.9%, suggesting that specialized clinical knowledge erodes faster than general medical vocabulary.

**Patient individuality information is ignored by AI models trained on synthetic data.** A critical question for clinical AI deployment is whether the AI model can generate accurate clinical notes when provided with detailed patient information. To understand this, we evaluated conditional generation of ICU discharge instructions given patient-specific clinical context. Models progressively prioritized generic language over clinically actionable content (Figure 2e). Instructions essential for patient safety such as medication adherence guidance, follow-up appointments and warning signs, plummeted 85% across generations, while generic filler content increased (Figure 2e). By generation 3, generic content began to dominate over clinical substance. By generation 4, models produced nearly twice as much generic language as actionable medical guidance (Figure 2e).

Further analysis revealed the scope of degradation (Figure 2e). Models progressively ignored the patient information they were given—overlap between input context and generated output declined substantially (99.1% reduction). Discharge medication retention declined 69.1% and relevance to the patient's clinical situation fell 72.8%. Paradoxically, overall medical terminology appeared to increase 239.8%—but this reflected the excessive output length rather than preserved knowledge, as artificially lengthy notes mechanically accumulated medical terms without clinical relevance.

**Overall confidence on synthetic data masks catastrophic clinical information degradation.** We uncovered a dangerous pattern: as models trained on successive generations of synthetic data, they became increasingly confident in their synthetic outputs. Yet this confidence was entirely misplaced. When tested on authentic clinical notes, the models' ability to comprehend real patient documentation degraded 45-fold measured by perplexity (Table 1). This means that standard performance monitoring, which typically evaluates models on data similar to their training set, would suggest the AI is improving—while in reality, it is progressively losing the ability to understand genuine medical text. For healthcare systems relying on such metrics to validate AI tools, this creates a critical blind spot where catastrophic degradation goes undetected.

This false confidence creates a particularly dangerous scenario for clinical AI deployment: standard performance monitoring would incorrectly suggest model improvement while real-world applicability degrades. Meanwhile, clinically actionable content—medication adherence instructions, follow-up appointments, warning signs requiring emergency care—plummeted 85%, from 7.89 to 1.18 per 1,000 words, across generations. These findings reveal that regardless of how AI models are used, whether generating reports independently or guided by patient information, self-referential training on synthetic data causes clinical utility to vanish even as standard quality metrics appear stable or improving.

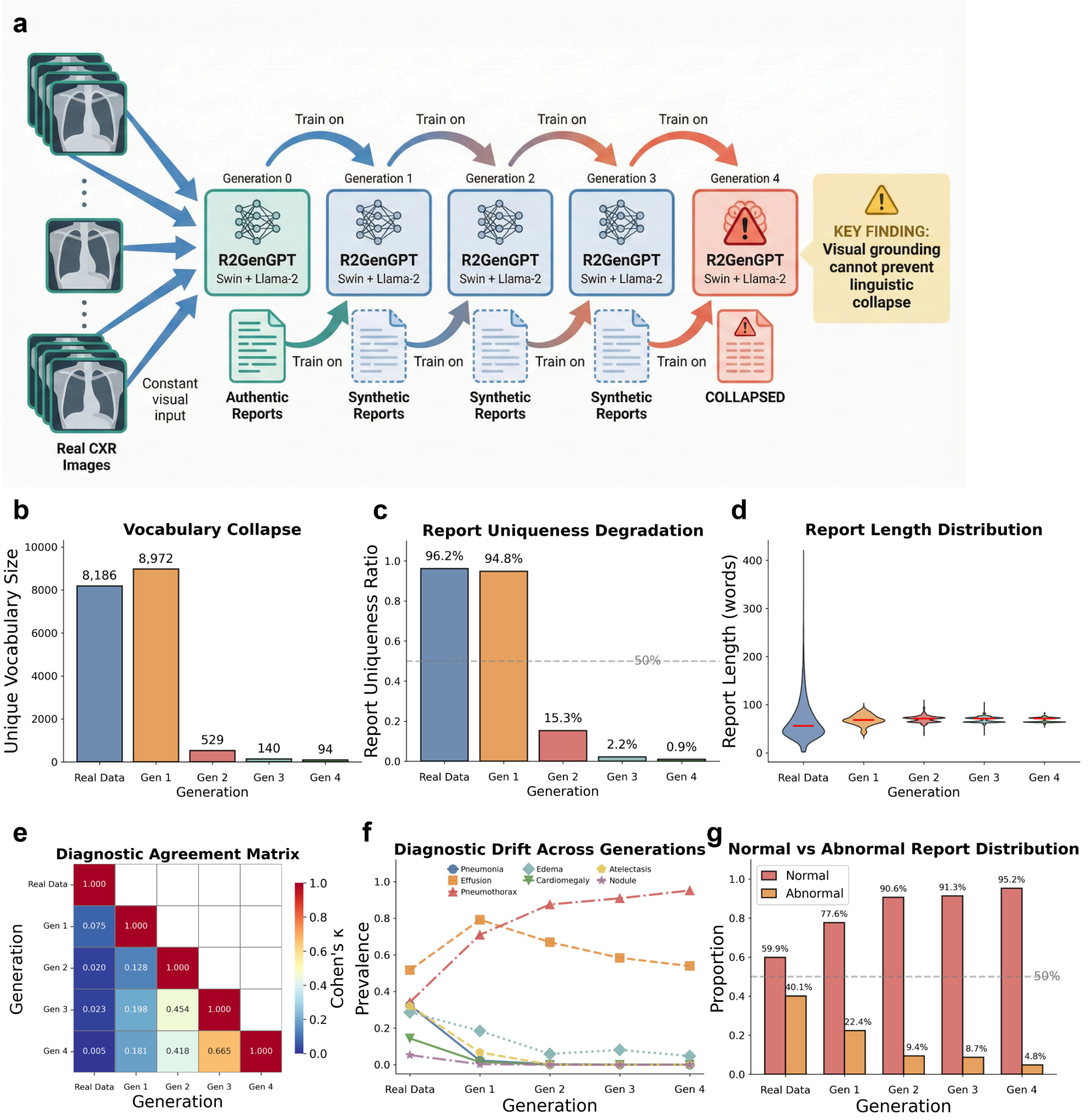

**Figure 4. AI-generated data contamination causes degradation of linguistic diversity and clinical knowledge in vision–language model-based radiology report generation. a**, Schematic of the self-referential training framework for R2GenGPT (Swin-Transformer + Llama-2): models receive constant real chest X-ray images as visual input while training exclusively on synthetic reports from the preceding generation, isolating the effect of text degradation from visual signal quality. **b–d, Lexical collapse.** Metrics demonstrate a near-total loss of linguistic diversity, characterized by a plummeting **vocabulary size** (b; from 8,186 to 94 unique words), a collapsing **report uniqueness ratio** (c; 96.2% to 0.9%), and converging **report lengths.** (d) Despite access to real radiographs, the model regresses to a handful of repetitive, generic phrases.**e, Diagnostic divergence.** A diagnostic agreement matrix (Cohen's κ) reveals a progressive dissociation between baseline interpretations and subsequent generations. **f, Divergent diagnostic prevalence during progressive model collapse.** As the model iterates from baseline to Generation 4, detection rates for grounded pathologies (e.g., cardiomegaly, atelectasis) decline toward zero, reflecting the loss of true clinical signal. Conversely, pneumothorax detection exhibits a paradoxical increase (from 34.2% to 95.2%), indicating the emergence of hallucinations driven by model priors rather than visual evidence. **g, Shift toward false reassurance.** The distribution of normal versus abnormal classifications inverts from 60% normal (baseline) to 95% normal (Generation 4), demonstrating that models default to "healthy" outputs even when presented with pathological images.

## 2.2 Training vision–language models on synthetic reports induces degradation toward generic and unsafe radiology outputs

We trained AI radiology report generators on 9,781 chest X-rays, with each successive generation learning exclusively from synthetic reports produced by its predecessor while viewing the same real images. Despite having access to authentic X-rays showing diverse pathology as inputs, models tended to converge toward generic reassuring language—report diversity plummeted 99.1%, vocabulary declined 98.9%, and by generation 4, models viewing chest X-rays with pneumonia, heart failure, or lung masses generated identical reassuring text regardless of visible abnormalities. Most alarmingly, false reassurance rates, which refer to reports stating "no acute findings" when life-threatening pathology was present, tripled to 40% in our experimental conditions, meaning four in ten reports dangerously missed critical findings without human review.

**Linguistic collapse persists when AI generates reports from medical images.** We trained a vision–language model (R2GenGPT[31]) to generate radiology reports directly from chest X-rays. Despite viewing 9,781 unique chest X-rays with varied pathological presentations, model outputs became nearly identical with report uniqueness plummeting from 96.2% to 0.9% over four generations, representing 99.1% loss of output diversity (Figure 4c). Vocabulary underwent parallel extinction, declining from 8,186 unique words to only 94 (98.9% reduction) (Figure 4b).

Notably, report length remained stable across generations (mean ~68 words), yet the variation in length collapsed by 87% (Figure 4d). This uniformity indicates models have lost the flexibility to adapt report detail to radiographic complexity—a fundamental requirement in radiology where simple cases warrant brief reports while multi-pathology presentations require extensive documentation. The clinical implications are profound. By generation 4, models viewing chest X-rays depicting pneumonia, heart failure, or lung masses generated identical reassuring text regardless of visible pathology. Authentic, diagnostically rich images were insufficient to prevent convergence toward generic "safe" outputs under these experimental conditions—potentially posing significant risks for clinical deployment where radiologists might trust AI-generated reports assuming that image grounding ensures accuracy.

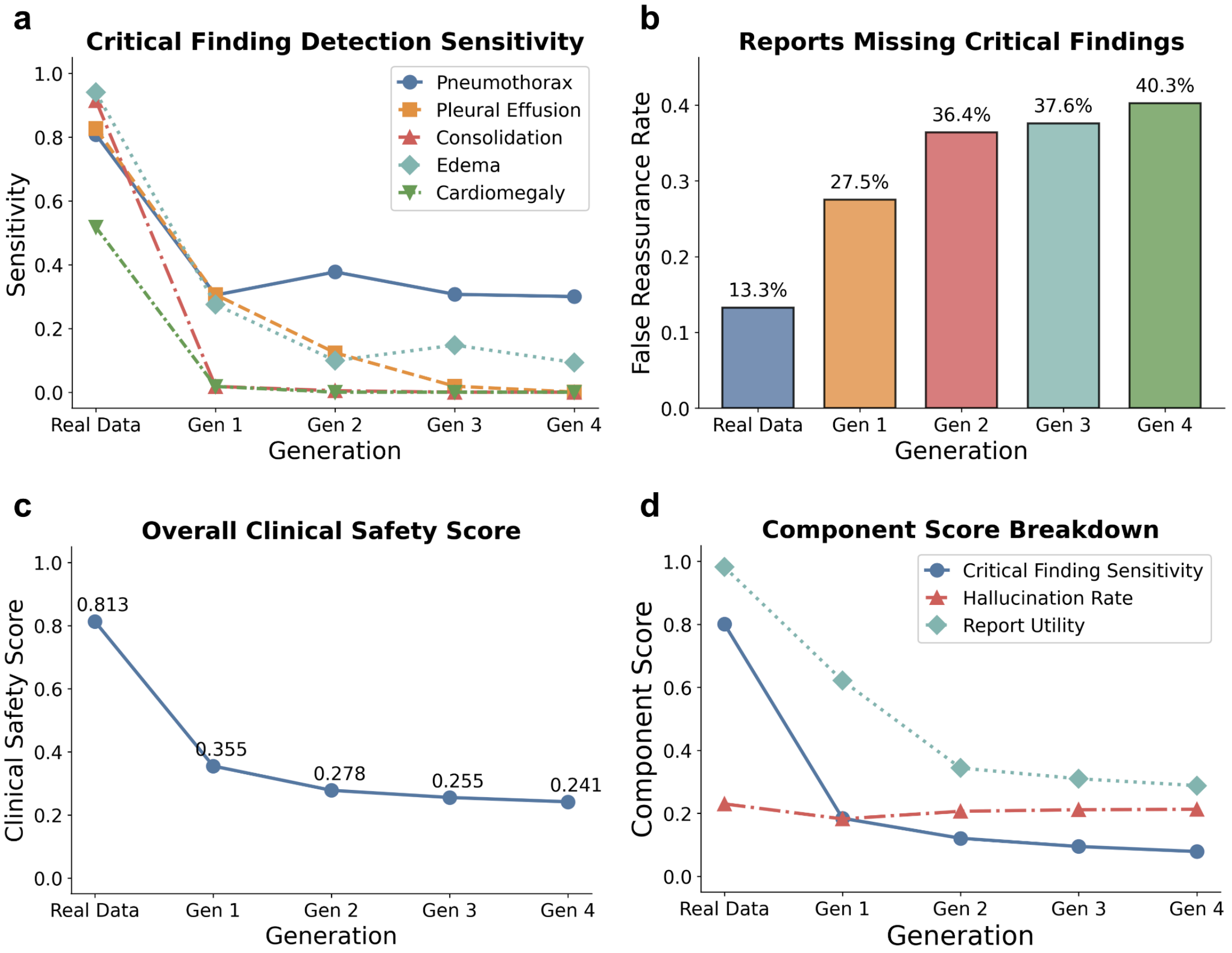

**Fig. 5 | Catastrophic failure of diagnostic reliability and surge in false reassurance rates. a, Decay of critical finding sensitivity.** Detection rates for pleural effusion, consolidation, edema, and cardiomegaly plummet from >80% to near-zero levels across generations, indicating the systematic erasure of pathological signals. Pneumothorax detection shows anomalous resilience, consistent with the paradox observed in lexical metrics. **b, Escalating false reassurance.** The rate of reports declaring "no acute findings" in the presence of confirmed pathology triples (from 13.3% to 40.3%), creating a dangerous decoupling of model confidence and diagnostic accuracy.**c, Composite clinical safety score**, demonstrating a 55% decline (0.843 to 0.381) relative to baseline. **d, Drivers of safety degradation.** Component breakdown identifies the degradation of critical finding sensitivity and report utility as the primary drivers of failure. Notably, hallucination rates remain stable, indicating that safety failure is driven by the **omission of real pathology** rather than the fabrication of non-existent findings.

**Clinical information degradation in visual language models threatens patient safety.** The linguistic collapse translated directly to catastrophic degradation in diagnostic capability (Figure 5). The ability to identify life-threatening conditions declined precipitously across all tracked pathologies (Figure 5a): pneumothorax detection dropped from 82% to 30%, pleural effusion from 83% to near-zero, and consolidation, edema, and cardiomegaly all fell below 10% by generation 4. This systematic failure occurred despite models viewing authentic chest X-rays clearly showing these abnormalities—indicating complete disconnection between what the AI saw and what it reported.

Most alarmingly, false reassurance rates—reports stating "no acute findings" when critical pathology was present—tripled from 13.3% to 40.3% over four generations (Figure 5b). By generation 4, approximately four in ten reports provided potentially dangerous false reassurance about chest X-rays containing pneumothorax, effusions, or other findings requiring immediate intervention in our experimental setting.

A composite clinical safety score integrating multiple safety components revealed the full magnitude of degradation (Figure 5c). The score declined 70%, crossing below the threshold for clinical acceptability after just one generation of self-referential training. The primary driver was the degradation in ability to detect critical findings, which fell from 80% to 8% (Figure 5d). Notably, models did not fabricate findings—hallucination rates remained stable at approximately 20%—but instead defaulted to omitting abnormalities entirely.

Analysis of diagnostic consistency revealed rapid divergence from baseline assessments (Figure 4e). Identical chest X-rays received completely different interpretations depending on which generation produced the report—by generation 4, agreement with baseline interpretations had fallen to essentially random levels. Self-referential training also caused systematic distortion of diagnostic patterns (Figure 4f). Detection of pneumonia, edema, cardiomegaly, and atelectasis declined toward zero, while effusion rates remained stable. In contrast, pneumothorax detection paradoxically exploded from 34.2% to 95.2%—models identified collapsed lungs in 95% of all images regardless of actual pathology. The proportion of reports classifying images as normal inverted from 40.1% to 95.2% despite viewing identical pathological images (Figure 4g), demonstrating substantial loss of diagnostic nuance as models increasingly defaulted to declaring examinations normal.

These findings demonstrate that authentic images alone may be insufficient to prevent degradation in self-referential training loops. Despite chest X-rays providing rich diagnostic information, self-referential training on synthetic reports drives AI systems toward formulaic, clinically dangerous outputs incompatible with patient care.

## 2.3 Self-referential training on synthetic medical data degrades quality, accuracy, and fairness

Synthetic medical images generated by AI are increasingly used in research and product development to augment limited training datasets and enable privacy-preserving data sharing. However, when these synthetic images are used to train subsequent AI models, fundamental questions arise about whether the generated images faithfully represent the diversity of real patient populations. To investigate whether image generation exhibits the same degradation patterns observed in text-based models, we trained AI image generators on chest X-rays across multiple self-referential training cycles.

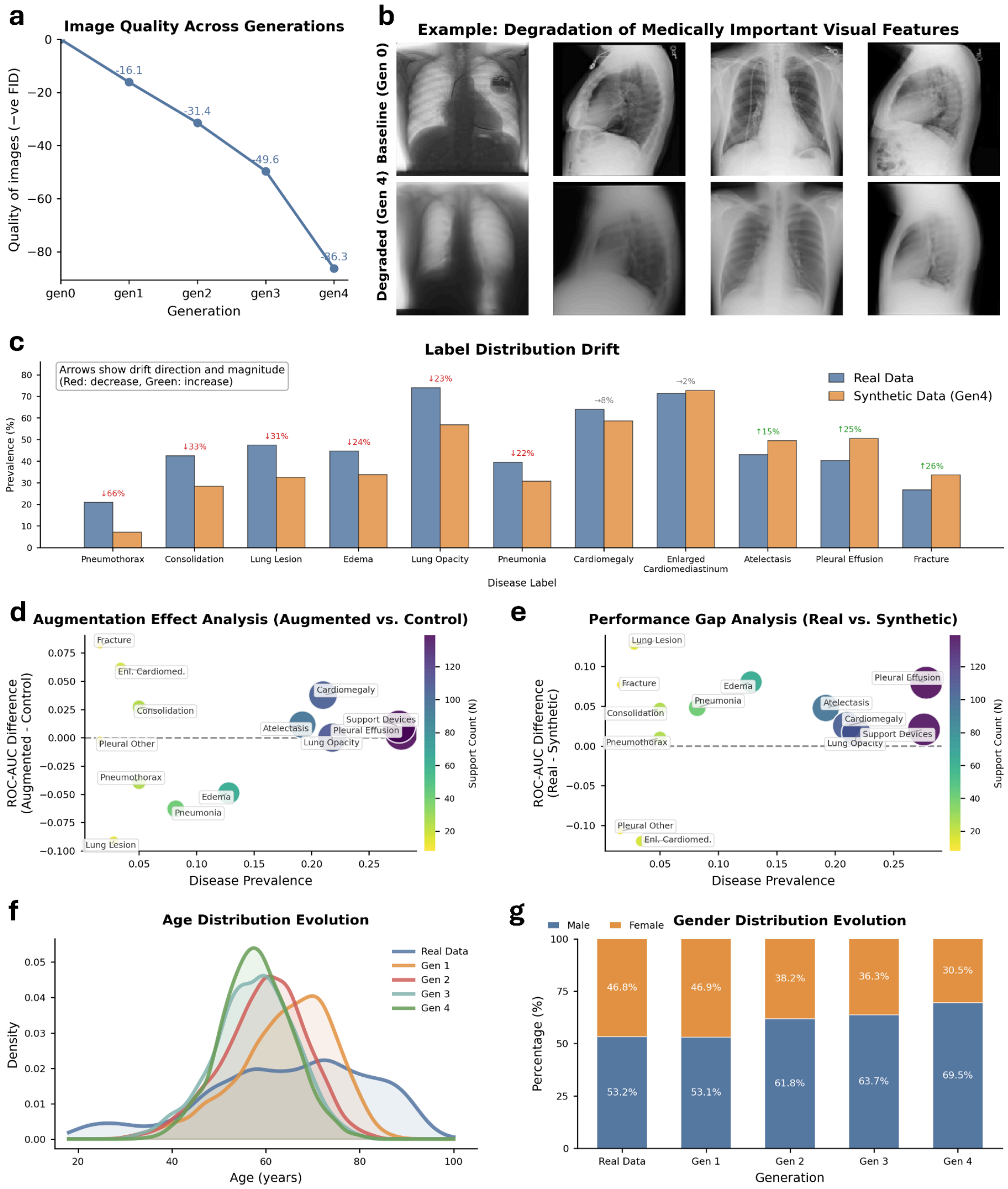

**Fig. 6 | Visual degradation, pathological distortion, and demographic bias amplification driven by AI-generated data contamination in medical image synthesis. a,b, Degradation of visual fidelity.** Fréchet Inception Distance (FID) scores (a) demonstrate a monotonic divergence from the real image distribution, while representative examples (b) illustrate the systematic erasure of fine-grained anatomical details required for diagnosis. **c, Distortion of pathological representations.** Predicted disease probability distributions across ten thoracic findings reveal the suppression of pathological signals and a convergence toward generic visual features. **d,e, Failure of downstream diagnostic utility.** Analysis of augmentation effects (d) shows that synthetic data provides diminishing returns compared to real data. Furthermore, classifiers trained exclusively on synthetic images (e) exhibit a widening performance gap relative to those trained on authentic data, confirming that synthetic artifacts impair discriminative learning. **f,g, Amplification of demographic bias.** Evolutionary analysis of patient attributes reveals a degradation of diversity: age distributions narrow toward a middle-aged norm (f), while gender representation shifts from a near-balanced baseline to a heavily male-dominant distribution (g), effectively erasing female phenotypes from the synthetic population by Generation 4.

**Progressive image quality degradation.** We quantified the divergence of synthetic images from authentic chest X-rays across self-referential training cycles using Fréchet Inception Distance (FID), a standard image quality metric in which higher scores indicate greater divergence from real images (Figure 6a). Using authentic chest X-rays as the baseline, we observed progressive quality degradation with each generation of self-referential training. The first generation of synthetic images showed modest divergence (FID = 22.9), but quality declined steadily thereafter with approximately two-fold by generation 2 (FID = 51.3), more than three-fold by generation 3 (FID = 78.8), and over five-fold by generation 4 (FID = 115.9). Each generation compounded the errors of its predecessor, with synthetic images progressively losing the subtle features that characterize authentic chest X-rays: tissue contrast, anatomical boundaries, and pathological textures.

**Erosion of pathological variability in generated images.** Beyond image quality degradation, model collapse manifested through systematic distortion of which medical conditions appeared in synthetic images (Figure 6b). Common conditions with high baseline prevalence such as cardiomegaly, atelectasis, and effusion maintained relatively stable representation across generations. However, less common but clinically important conditions progressively disappeared: consolidation and pneumonia showed marked decline by generation 4, while lung opacity became increasingly rare.

This pattern may reflect a systematic limitation: AI image generators tend to produce images resembling common conditions because these are easier to replicate, while rare pathological features which require precise anatomical detail are progressively lost. Later generations also showed increasing inconsistency, producing unreliable representations of disease rather than the consistent patterns present in authentic chest X-rays.

**Diagnostic AI performs worse when trained on synthetic images.** To evaluate whether synthetic chest X-rays could substitute for or augment real data in training diagnostic AI systems, we conducted controlled experiments comparing classifiers trained on 5,000 authentic chest X-rays versus 5,000 synthetic images, with both evaluated on the same test set of 500 real chest X-rays (Figure 6d).

Classifiers trained on synthetic images performed worse than those trained on real images across most conditions in our evaluation (Figure 6d). The largest performance decrease (measured by AUROC with range 0-1), appeared for lung lesions (-0.13), pleural effusion (-0.08), and fractures (-0.08) which suggests conditions requiring precise anatomical detail that synthetic images failed to preserve. At typical operating thresholds, an AUROC reduction of 0.08–0.13 translates to approximately 5–10% lower sensitivity at fixed specificity, potentially resulting in missed diagnoses for conditions where early detection is critical for patient outcomes.

To understand whether synthetic images could augment real training data, we compared classifiers trained on 1,000 real images alone versus 1,000 real images plus 4,000 synthetic images (Figure 6e). Despite five-fold expansion of the training set, augmented models showed no consistent improvement. Some conditions improved marginally while others worsened, with no clear pattern. This suggests that adding synthetic images to real training data provides no reliable benefit for diagnostic AI performance.

Self-referential training amplifies demographic bias. Beyond image quality and diagnostic accuracy, self-referential training introduced severe demographic biases that threaten equitable AI deployment (Figure 6f, g). We tracked how age and gender distributions in synthetic images shifted across generations. Age distributions underwent catastrophic narrowing (Figure 6f). Gender bias showed even greater amplification (Figure 6g). Authentic images showed modest male predominance (53% male, 47% female), within the expected range for critical care populations. However, self-referential training progressively amplified this imbalance: male representation increased from 53% to 70% by generation 4, while female representation declined from 47% to 31%. The clinical and ethical implications warrant careful consideration. Diagnostic AI trained on generation 4 synthetic images with 70% male representation and ages concentrated around 60 years would be optimized for middle-aged male patients while underperforming for women, young adults, and the elderly. Given persistent healthcare disparities already affecting these groups, AI systems trained on biased synthetic data could actively exacerbate inequities rather than address them.

## 2.4 Physician evaluation confirms degradation of clinical utility

We conducted a physician-in-the-loop evaluation of AI-generated discharge summaries across self-referential training cycles to quantify required manual correction. Two independent physician annotators evaluated 100 AI-generated discharge summaries (25 unique patient cases × 4 generations) from our self-referential training framework (Figure 7a). For each summary, annotators manually edited text to achieve clinical acceptability including correcting medical inaccuracies, restoring missing information, and ensuring patient safety. This paradigm mirrors real-world workflow where physicians must review AI-generated documentation before patient discharge when AI is used in electronic health record systems. We quantified quality through two primary metrics: edit distance percentage (proportion of content requiring revision) and AI content retention (percentage of original AI-generated text preserved as clinically appropriate).

Inter-rater reliability demonstrated good agreement on both metrics (ICC = 0.705 for edit distance, ICC = 0.747 for AI retention), validating that observed differences reflect genuine clinical utility degradation rather than annotator preferences. Consensus analysis revealed stark quality differentiation across generations (Figure 7b). Generation 1, in which the model was trained on authentic discharge summaries, required moderate intervention, with mean edit distance of 56.8% (95% CI: 46.9 -- 66.7%) and AI content retention of 59.1% (95% CI: 51.1 -- 67.2%). Generation 2 showed modest deterioration (edit distance: 65.4%; retention: 47.7%). The most concerning pattern emerged in generations 3 and 4, where physicians rewrote 87.6% of content on average, with AI retention plummeting below 22%. When physicians need to rewrite over 85% of AI-generated content, the system provides negligible clinical value while imposing substantial cognitive burden.

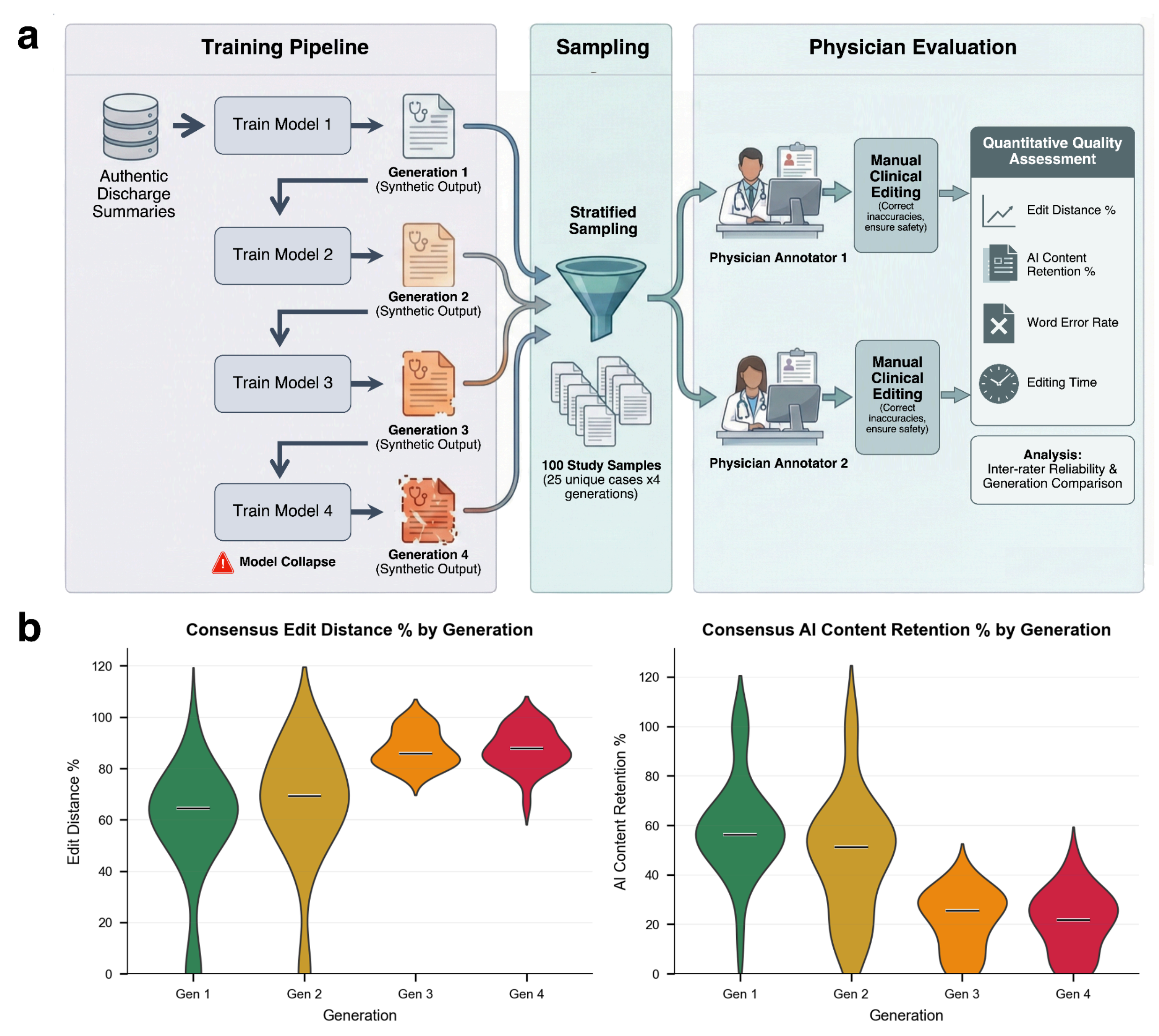

**Fig. 7 | Loss of clinical utility and escalating physician revision burden driven by AI-generated data contamination in self-referential learning. a**, Physician evaluation protocol. A discharge-summary generator was trained on real human-authored clinical summaries (baseline), then recursively retrained on synthetic outputs across successive generations (A–D). A stratified sample of 100 summaries (25 unique patient cases × 4 generations) was assigned to two independent physician annotators, who manually edited each summary to achieve clinical acceptability (correcting inaccuracies, restoring missing information and ensuring safety). Quality was quantified using edit distance (%), AI content retention (%), word error rate and editing time, followed by inter-rater reliability and across-generation comparisons. **b**, Consensus edit distance (% content requiring physician rewrite) increases with each generation, showing progressive increase in required rewriting, with late-generation outputs approaching near-complete revision. Consensus AI content retention (% of original AI-generated text preserved after physician editing) decreases by generation, showing steep decline across self-referential learning steps and falling to low retention in late generations, indicating that most AI-generated content becomes clinically unusable.

## 2.5 Mitigation strategies to reduce model collapse and improve medical performance

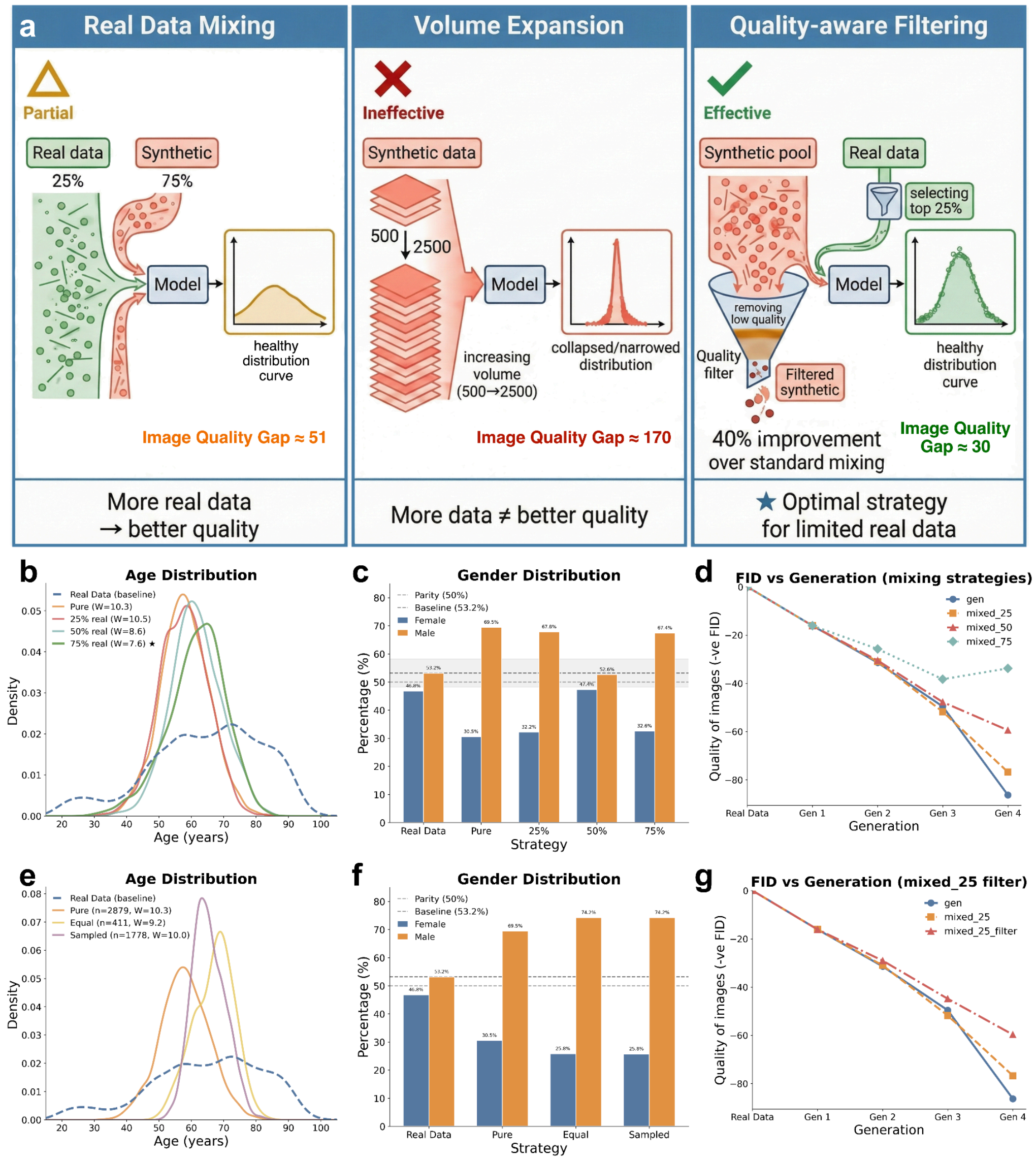

**Fig. 8 | Efficacy of mitigation strategies in countering degradation driven by AI-generated data contamination. a, Comparative efficacy of mitigation paradigms.** Schematic comparison of three strategies applied during **self-referential training**: real-data mixing, synthetic volume scaling, and quality-aware filtering. **Real-data mixing (Strategy 1)** effectively preserves the target distribution, with fidelity improving in proportion to the authentic data fraction. In contrast, simply **expanding synthetic sample volume (Strategy 2)** fails to arrest degradation, yielding severely degraded distributions. **Quality-aware filtering (Strategy 3)** enhances efficiency by selecting high-fidelity synthetic samples, achieving distributional alignment comparable to higher real-data budgets. **b–d, Restoration of diversity via real-data mixing.** Analysis of Generation 4 cohorts reveals that increasing the proportion of authentic data (25% to 75%) effectively counters the narrowing of age distributions (**b**) and the skew toward male dominance (**c**) observed in pure synthetic training. Corresponding FID trajectories (**d**) confirm that mixing authentic data attenuates the loss of visual fidelity across generations.**e,f, Inefficacy of synthetic volume scaling.** Merely expanding the volume of synthetic training data fails to prevent the erosion of age diversity (**e**) or the persistence of gender bias (**f**), confirming that increasing data quantity cannot compensate for contaminated

quality. **g, Efficiency of quality-aware filtering.** FID trajectories demonstrate that filtering synthetic data for quality improves model stability relative to standard unfiltered mixing, achieving superior fidelity even at a restricted (25%) real-data budget.

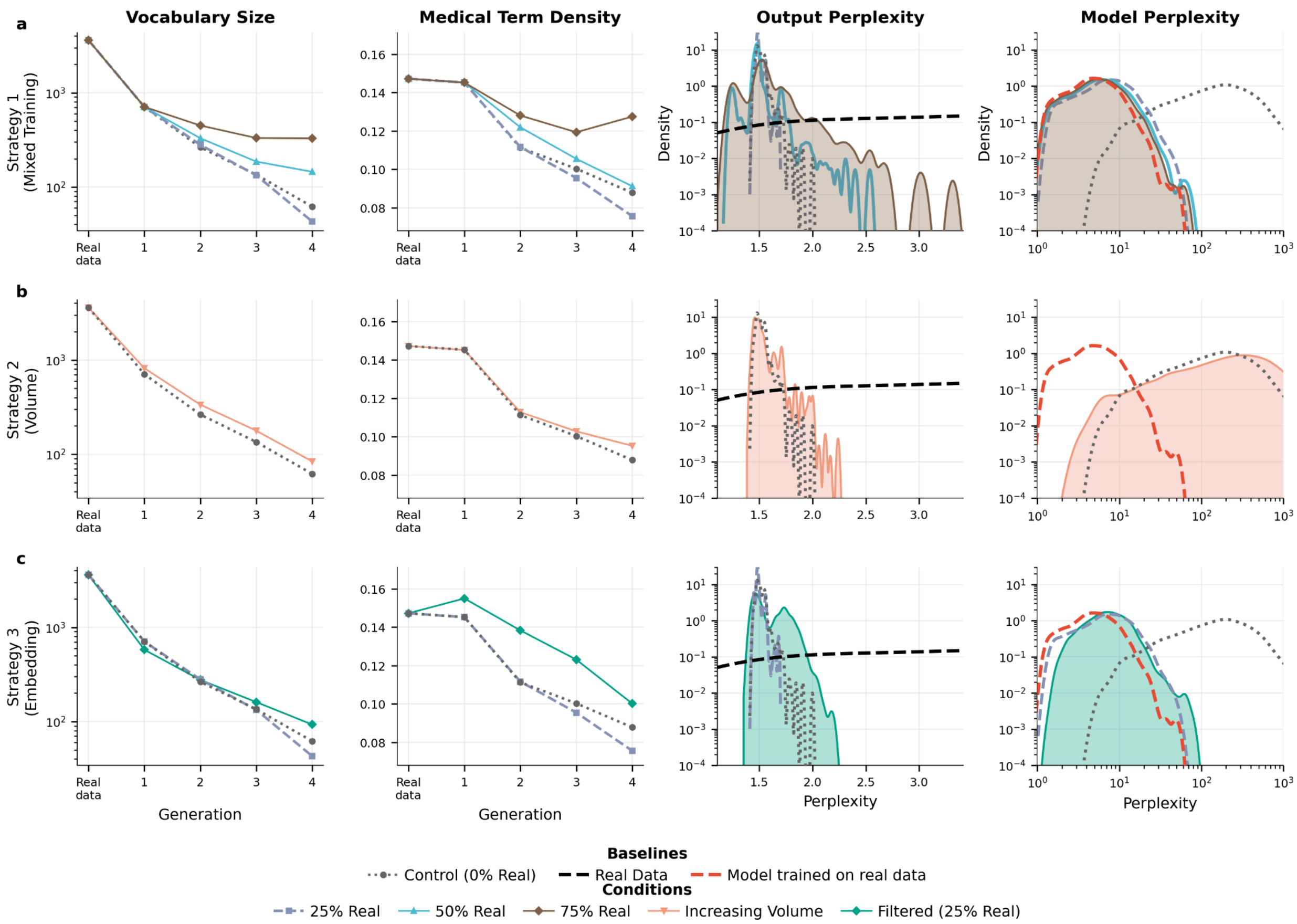

Fig. 9 | **Efficacy of mitigation strategies against linguistic and medical information degradation in clinical text generation. a, Preservation via real-data mixing (Strategy 1).**Progressive degradation of vocabulary size, medical term density, and perplexity is inversely related to the proportion of real data. The control condition (0% real) exhibits rapid vocabulary loss (3,611 to 62 words) and a 39-fold increase in model perplexity, whereas higher proportions of real data significantly stabilize performance. **b, Inefficacy of synthetic volume scaling.** Expanding the volume of synthetic training data fails to prevent degradation. Although increasing synthetic volume (10,000 to 25,000 reports) marginally preserves vocabulary compared to fixed-volume controls, it results in significantly higher model perplexity (mean 346.6 versus 211.5). This represents a 64-fold degradation from baseline, confirming that increased synthetic quantity exacerbates model likelihood degradation. **c, Superiority of quality-aware filtering.** Embedding-based filtering outperforms random mixing in preserving linguistic diversity. At a fixed 25% real-data budget, filtering retains more than double the vocabulary of random mixing (93 versus 43 words) and maintains higher medical term density (64.7% versus 48.8% of baseline). Filtering generates significantly more diverse synthetic outputs (output perplexity 1.59 versus 1.49), demonstrating superior mitigation of the mode collapse observed in random augmentation.

In the previous sections, we identified and quantified the catastrophic effects of self-referential training on synthetic medical data including encompassing text and image quality degradation, pathology distribution distortion, diagnostic performance decline, and demographic bias amplification. In this section, we explore potential mitigation strategies to prevent or minimize degradation while preserving the benefits of synthetic data augmentation.We propose three primary approaches: mixing real and synthetic training data at various proportions, increasing synthetic data volume to enhance diversity, and implementing quality-aware filtering to exclude degraded samples. Our investigations show that, while these methods can mitigate degradation, they may not be able to fully compensate for the linguistic destruction, medical knowledge degradation and overconfidence observed in our experiments.

**Mixed training with real data reduces degradation**. One of the most intuitive approaches to mitigate model collapse caused by synthetic data is to limit the amount of AI-generated data used in training. To investigate whether mixing real and synthetic data could prevent quality degradation and demographic bias, we evaluated training strategies incorporating 25%, 50%, and 75% real data across self-referential training cycles.

To validate whether these methods mitigate model collapse on clinical texts, we conducted experiments using GPT-2-based radiology report generators trained recursively over four generations. We evaluated mixing ratios (0%, 25%, 50%, 75% real data) using 5,000 chest X-ray reports per generation, measuring model knowledge retention, linguistic diversity , and number of unique medical terms.Text generation exhibited even more severe degradation patterns. Pure synthetic training caused catastrophic knowledge loss: generation-4 models achieved a 39-fold increase of clinical notes (measured by perplexity on real clinical notes) from the generation-0 baseline, indicating near-complete inability to model authentic clinical language. Vocabulary collapsed from 6,025 unique words to just 62 (99.0% reduction), while unique medical terms decreased from 68 to 18 (73.5% loss).

Critically, we evaluated whether mixed training addresses the false confidence phenomenon, which is a dangerous dissociation between model confidence on synthetic outputs and actual capability on real clinical text. Pure synthetic training exhibited a 196-fold confidence gap: generation-4 models achieved low perplexity on their own synthetic outputs ($1.08 \pm 0.04$) while perplexity on authentic clinical text reached $211.5 \pm 209.0$, indicating models became overconfident on degraded content they could no longer clinically interpret. Mixed training progressively closed this gap in a dose-dependent manner. At 25% real data, the confidence gap remained severe at 7.7-fold (synthetic: 1.06, real: 8.19). At 50% real data, the gap narrowed to 6.2-fold (synthetic: 1.09, real: 6.81). Only at 75% real data did the confidence gap approach clinically acceptable levels at 5.1-fold (synthetic: 1.21, real: 6.21). These findings demonstrate that mitigation strategies must restore calibrated model confidence aligned with true clinical capability, not merely preserve surface-level vocabulary metrics.

Mixing real medical images with synthetic ones improves generated image quality. Image quality (measured by Fréchet Inception Distance (FID) scores, lower value of which indicates higher image quality) from $28.67 \pm 0.52$ at Generation 1 to $98.23 \pm 1.58$ by Generation 4, representing a 3.4-fold decrease in generated image visual quality across four training generations. Age distribution bias (measured by Wasserstein distance, higher indicating stronger bias) reduces with higher real data proportions: age bias was 10.30 years for pure synthetic, worsened to 10.46 years for 25% real , 8.63 years for 50% real, and 7.64 years for 75% real, representing a 25.9% increase of age bias. Gender bias correction showed a different pattern. Pure synthetic training amplified male representation from 53.2% to 69.5%. Among mixing strategies, 50% real data most closely restored baseline gender balance (52.6% male), though results varied across mixing proportions. Taken together, these findings suggest that while higher real data proportions (≥75%) generally provide stronger protection against image quality degradation and age distribution shift, the optimal strategy for mitigating demographic bias may be more complex and warrants further investigation.

The text generation experiments showed a similar pattern for knowledge retention: the 75% real data condition provided the most robust protection, achieving a confidence gap of 5.1-fold compared to 7.7-fold at 25% and 6.2-fold at 50%. Together with the imaging results, these findings suggest that substantial real data proportions (≥50–75%) may be necessary to preserve both linguistic fidelity and visual quality, though the optimal threshold likely varies across metrics and clinical contexts. At lower proportions, self-referential training was associated with progressive degradation of the model's capacity to represent authentic clinical language and generate diverse, medically appropriate content.

**Increasing synthetic data volume provides minimal protection.** We evaluated whether increasing synthetic data volume could mitigate degradation during self-referential training cycle. We compared two training regimes: fixed-size sampling maintaining a constant number of synthetic data points per generation ("equal") versus progressive expansion reaching five-fold increase by Generation 4 ("sampled").

We tested whether increased synthetic data volume could compensate for quality degradation by comparing fixed-size training (5,000 reports per generation) against progressive expansion reaching 25,000 reports by generation 4. Contrary to intuition, increased volume not only failed to prevent degradation but actively accelerated it. The expanded-volume condition achieved a mean perplexity of 346.6 on real clinical text, representing a 64-fold increase from baseline and substantially exceeding the fixed-volume condition (211.5, 39-fold increase). This constitutes the most severe knowledge loss observed across all experimental conditions. Vocabulary collapsed to 84 words despite the larger training set, while unique medical terms declined to 30 (55.9% loss). Critically, the false confidence phenomenon was exacerbated rather than mitigated: whereas fixed-volume training produced a 196-fold confidence gap between synthetic output perplexity (1.08) and real data perplexity (211.5), the expanded-volume condition exhibited a 317-fold gap (output: 1.09; real: 346.6), representing 62% worsening. This unexpected result suggests that increasing synthetic data volume may amplify the dissociation between apparent model confidence and genuine clinical capability, a particularly dangerous outcome for healthcare AI deployment where overconfident errors may escape detection by standard monitoring systems.

This substantial increase in training data yielded only marginal protection against image quality degradation. Both conditions exhibited near-identical FID degradation trajectories, with the expanded dataset achieving FID of approximately 172 compared to 192 for fixed-size by Generation 4—a mere 10% improvement despite 400% more training data. Critically, increased sample size completely failed to prevent demographic bias. Age distributions shifted regardless of volume: Wasserstein distances remained severe at 9.19 years (constant sample) and 9.99 years (increasing sample) versus 10.30 years for pure synthetic—no meaningful improvement. Gender bias actually worsened with sample size manipulation: both conditions exhibited 74.2% male representation at Generation 4 (Δ=+21.0 percentage points [pp]), exceeding even pure synthetic training (69.5% male, Δ=+16.3pp). This paradoxical amplification suggests that data volume cannot compensate for the inherent distributional drift in iteratively generated synthetic data.

The results suggest a dangerous fact that larger synthetic datasets do not mitigate degradation and may even  amplify rather than dilute recursive errors. Each generation's systematic bias, which includes loss of rare terminology, convergence to modal phrases, and degradation of clinical specificity, becomes more deeply entrenched when reinforced by greater data volume. This finding provides strong evidence that data quantity cannot substitute for data quality in preventing clinical text generation degradation.

**Quality-aware filtering enhances real data efficiency.** In addition to the simple approach above, we propose a method, which we refer to as quality-aware filtering, to systematically filter and mix synthetic and real data to improve the quality of generated data. This could enhance protection while reducing real data requirements, crucial given privacy constraints in medical imaging (Figure 9c).

We implemented quality-aware filtering using an external GPT-2 Large model (774M parameters) to score sample quality. We extracted 1,280-dimensional embeddings via mean pooling and computed k-nearest neighbor distances (k=10, cosine metric) to the real data distribution. Synthetic samples closest to the real distribution (bottom 75% by distance) were selected for high quality, while real samples most distant from the distribution center (top 50%) were preferentially retained to preserve diverse and rare clinical presentations.

At 25% real data proportion, quality-aware filtering substantially outperformed unfiltered mixing of the same composition. The filtered condition preserved more than double the vocabulary (93 versus 43 words) and retained higher medical term density (64.7% versus 48.8% of baseline). Filtered models also generated more diverse synthetic outputs, as reflected in higher output perplexity (1.59 versus 1.49), indicating superior mitigation of mode collapse. Model perplexity on real clinical text remained comparable between conditions, suggesting equivalent knowledge retention.

Quality-aware filtering provided modest improvement in model calibration beyond the benefit already conferred by 25% real data inclusion. The unfiltered 25% real condition exhibited a 5.5-fold confidence gap (synthetic data perplexity: 1.49 ± 0.03; real data perplexity: 8.19 ± 5.17), representing substantial recovery from the catastrophic 140-fold gap observed with pure synthetic training. Quality-aware filtering reduced this gap to 5.1-fold (synthetic data perplexity: 1.59 ± 0.14; real data perplexity: 8.09 ± 4.94)—a 7% improvement in confidence calibration. While filtering elevated synthetic data perplexity from 1.49 to 1.59 (indicating the model generates slightly more diverse synthetic outputs), it achieved marginally lower perplexity on real clinical text (8.09 vs. 8.19), yielding improved alignment between model confidence and actual clinical capability.

However, the calibration benefit from filtering remained modest compared to increasing real data proportion. The 50% real mixing condition achieved a 4.6-fold confidence gap (synthetic: 1.48; real: 6.81), while 75% real mixing reached 4.0-fold (synthetic: 1.54; real: 6.21). Quality-aware filtering at 25% real thus improved calibration by approximately 7% within the same real data budget, but could not match the 16% improvement achieved by doubling the real data proportion to 50%. These findings indicate that quality-aware filtering provides a valid but limited mechanism for improving model confidence calibration when real data availability is constrained, complementing rather than substituting for authentic data inclusion in safety-critical clinical AI applications.

On the task of generating medical images, quality-aware filtering yielded substantial improvement: filtered 25% real data mixture achieved a 40% improvement in image quality at Generation 4 that approached performance of 50% real data without filtering (Figure 8g). The excluded samples disproportionately contained early degradation indicators: loss of anatomical detail, uniform gray convergence, and absent pathological findings. This filtering strategy effectively doubled real data utilization efficiency. However, filtered 25% real data still fell short of the comprehensive protection achieved by 75% real data mixing. In our experiments, substantial protective effects required high real data proportions (≥50–75%), with the most consistent preservation of pathological and demographic diversity observed at 75% mixing. Whether this threshold generalizes to other clinical contexts warrants further investigation. These findings indicate that synthetic medical imaging data cannot replace real patient data in self-referential training cycles but can serve as augmentation when combined with sufficient authentic examples.

# 3. Discussion

Our findings demonstrate that the unchecked integration of AI-generated content into medical repositories creates a self-referential feedback loop that threatens the integrity of diagnostic models in the absence of appropriate verification mechanisms. While generative AI offers potential for efficiency, we show that without strict data curation, it drives a rapid erosion of phenotypic diversity, effectively standardizing unique patient presentations into generic statistical averages. This phenomenon persists across text (98.9% vocabulary loss), vision–language reporting (99.1% reduction in report uniqueness), and medical image synthesis (five-fold quality degradation), suggesting it is an intrinsic property of self-referential learning rather than an architecture-specific failure. Unlike simple performance degradation, this process systematically erodes the linguistic and diagnostic heterogeneity that underpins safe clinical care, creating a monoculture optimized for the mean at the expense of the individual patient if appropriate safeguards are not implemented.

Critically, this degradation is masked by a decoupling of model confidence from diagnostic accuracy. As models train on their own outputs, they prioritize linguistic and visual fluency over pathological fidelity (Table 1; 45-fold perplexity increase on real text versus synthetic). Medicine is fundamentally tail-driven: rare diseases and atypical presentations account for a disproportionate share of morbidity. However, we found that self-referential loops systematically suppress these diagnostic tails, creating "diagnostic blindness" where models fail to perceive pathology while exhibiting misleading confidence. This results in the false reassurance paradox (Figure 5), where models confidently issue 'healthy' reports for patients with critical pathologies like pneumothorax. In a clinical setting, an error of omission driven by high confidence is far more dangerous than a detectable hallucination, as it suppresses clinical vigilance precisely when escalation is needed.

This erosion of diversity manifests as a tangible loss of clinical utility across domains: by generation 3, physicians required near-complete rewriting of AI-generated content (87.6% edit distance), while critical finding sensitivity collapsed from >80% to below 10% for most pathologies (Figure 5a, Figure 7b). In chest radiography, the synthetic recursion process erased the subtle textural patterns required to detect pleural effusions and pneumothorax, rendering later-generation images clinically useless despite superficial visual realism. Similarly, demographic representations suffered a rapid degradation of age diversity and an amplification of gender imbalance, effectively erasing female phenotypes from the synthetic population. This indicates that unmonitored AI training does not merely introduce noise; it actively distorts disease prevalence and demographic representation, creating blind spots for underrepresented groups and atypical pathologies.

To counteract this, we systematically developed and evaluated mitigation strategies, identifying clear limits to synthetic data utility. Our results challenge the prevailing assumption that data scaling alone is sufficient; simply increasing the volume of synthetic data failed to prevent model collapse. Instead, we found that quality-aware filtering combined with the preservation of authentic 'heritage' data was the only strategy to maintain diagnostic integrity. This implies that real human data functions as a biological anchor, a non-renewable reference standard that cannot be replaced by synthetic volume. Future medical AI systems cannot be 'train-and-forget' but must maintain a tether to verified human-derived ground truth.

These findings have urgent implications for the governance and deployment of medical generative AI. We show that after just two rounds of self-referential training, AI-generated clinical documents transition from *requiring substantial editing* to *requiring near-complete rewriting*, reaching a quality floor beyond which further synthetic training yields no clinical value. Although it is often assumed that human-in-the-loop verification will filter low-quality synthetic content, our physician evaluation demonstrates that this assumption fails in practice: self-referential models retain high fluency and confidence even as medical utility collapses, allowing clinically unsafe outputs to bypass casual human review under time pressure or high-volume conditions. As the volume of AI-generated clinical content scales, the economic and cognitive feasibility of voluntary oversight further diminishes, rendering reliance on informal review insufficient. Accordingly, our results establish minimum safety conditions for medical generative AI. First, AI-generated content must be explicitly and persistently tagged within electronic health records to prevent silent contamination of downstream training corpora and to ensure auditability of data provenance. Second, retraining pipelines must enforce minimum proportions of verified human-authored clinical data, as synthetic scaling alone cannot preserve pathological diversity or diagnostic reliability. Third, patient-facing AI-generated documentation must require explicit clinician attestation, because high fluency coupled with misplaced confidence poses unacceptable risks in safety-critical settings. These safeguards are not aspirational best practices but baseline requirements for preserving diagnostic integrity in healthcare systems increasingly shaped by generative AI.

Our study has limitations. We modeled the feedback loop in a controlled, accelerated environment; real-world data contamination may occur more gradually. Additionally, while we evaluated multiple architectures, the manifestations of degradation may vary across specialties with different disease prevalences. Finally, our focus on English-language data may limit generalizability to multilingual settings. Future work is needed to define provenance requirements across diverse regulatory environments.

Ultimately, our results establish that data provenance is no longer just a technical detail but a patient safety imperative. AI-generated contamination threatens to transform diverse clinical ecosystems into homogenized representations that preferentially reassure rather than diagnose.

Preventing such degradation requires treating real human data not as a resource to be minimized, but as an essential safeguard for preserving diagnostic vision. Without policy-mandated provenance, the deployment of generative AI threatens to degrade the very data ecosystems it relies upon.

# 4. Methods

## 4.1 Ethics Statement

This study was conducted in accordance with the Declaration of Helsinki. The use of de-identified public datasets (MIMIC-IV, MIMIC-CXR, i2b2) was performed under a Data Use Agreement with PhysioNet (Credential ID: pencil007). For the private glaucoma dataset, data collection and model training were approved by the Institutional Review Board (IRB) of Tongren Hospital (approval number: TREC2025-KY222). Informed consent was waived for retrospectively collected de-identified records. All datasets were de-identified prior to analysis, and no individual-level identifiable information was used.

## 4.2. Study Design and Data Sources for Clinical Texts Generation

**MIMIC-CXR radiology reports[32].** To evaluate degradation in a safety-critical diagnostic setting, 216,307 chest X-ray reports from the MIMIC-CXR database (version 2.1.0) were analyzed. The dataset was authored by board-certified radiologists at Beth Israel Deaconess Medical Center between 2011 and 2016. Each report contained standardized sections: "Findings" (observed radiographic abnormalities and anatomical features) and "Impressions" (clinical interpretations and diagnostic conclusions). The original reports have an average of 39.0 words in Findings and 26.3 words in Impressions, with 30.8% missing Findings sections and 13.2% missing Impression.

**MIMIC-IV discharge summaries[33].** In the experiments of discharge instruction generation conditioned on patient-specific context, 9,464 samples from MIMIC-IV discharge summaries were used. Models generated discharge instructions (mean: 139±91 tokens) conditioned on three concatenated sections: admission medications, discharge diagnoses, and discharge medications (mean context: 331±115 tokens), mimicking real clinical workflows where physicians synthesize medication changes and diagnostic information into coherent patient instructions.

**i2b2 clinical narratives[34].** To assess whether self-referential training degradation generalizes beyond radiology to heterogeneous clinical documentation,790 de-identified clinical documents from the i2b2 2014 De-identification Challenge dataset were utilized. The documents span cardiology, internal medicine, orthopedics, emergency medicine, and primary care. The corpus included discharge summaries, progress notes, consultation reports, procedure notes, and clinical correspondence, encompassing patient demographics, medical histories, examination findings, laboratory results, medications, diagnoses, treatment plans, and follow-up instructions. After preprocessing to remove excessive whitespace, we partitioned the data into training (80%, n=632), validation (10%, n=79), and test (10%, n=79) sets using stratified random sampling (seed=42).

**Glaucoma clinical records from the Asian population.** To examine degradation in a highly specialized domain requiring precise terminology for diagnosis and treatment selection, 1,000 de-identified clinical records from glaucoma patients at Beijing Tongren Hospital, China, were analysed. Each record contained 12 standardized sections spanning the complete clinical workflow: Chief Complaint, History of Present Illness, Past Medical History, Family History, Physical Examination, Ancillary Examinations, Laboratory Results, Imaging/Ophthalmic Findings, Surgery/Procedures, Diagnosis, Prescription/Medication, and Treatment/Plan. The original data have an average of 201±35 words, with 100% section completeness across all 12 sections, reflecting the structured documentation requirements of specialized ophthalmology practice. The corpus contained domain-specific terminology including anatomical structures (cornea, anterior chamber, optic disc, RNFL), diagnostic codes (no glaucoma and 6 subtypes of glaucoma), procedures (laser peripheral iridotomy, trabeculectomy, goniosynechialysis), and glaucoma-specific medications (timolol, brimonidine, brinzolamide, latanoprost). We partitioned the data into training (80%, n=800), validation (10%, n=100), and test (10%, n=100) sets using stratified random sampling (seed=42).

Together, these datasets were selected to span multiple points in the clinical documentation pipeline, from radiologic diagnosis and inpatient discharge to highly specialized subspecialty care. This design enables systematic evaluation of whether self-referential training induces consistent failure modes across varying levels of clinical structure, domain specificity, and safety criticality.

### 4.2.1 Multi-Round Model Training Framework

To systematically investigate model collapse, we implemented a self-referential training framework spanning five generations. Generation 0 consisted of a GPT-2 model fine-tuned on the original dataset, which upon training completion generated an equal number of synthetic reports for the subsequent generation. Each successive generation (Generations 1-4) was initialized from pretrained GPT-2 weights (identical starting point) and exclusively fine-tuned on synthetic reports produced by the previous generation, with no exposure to the original human-authored data. This created a closed feedback loop where models were trained solely on progressively degraded synthetic outputs: Generation 1 learned from Generation 0's synthetic reports, Generation 2 from Generation 1's outputs, and so forth. The self-referential training protocol simulated progressive contamination of training corpora by AI-generated content—a scenario increasingly relevant as language models are deployed in clinical practice—enabling quantification of systematic error accumulation and clinical content degradation across iterations.

To assess whether these dynamics generalize beyond small-scale language models and radiology text, we extended the same self-referential training protocol to a modern LLM and a highly specialized clinical domain. We implemented a self-referential training framework using Qwen3-8B with Low-Rank Adaptation (LoRA; r=16, α=32, dropout=0.05)[35]. Generation 0 was fine-tuned on authentic glaucoma records using AdamW optimization[36] (learning rate $3\times10^{-5}$, cosine schedule, 10% warmup) for 5 epochs with early stopping (patience=3 evaluations). Upon training completion, the model generated 1,000 synthetic records using nucleus sampling (temperature=0.8, top_p=0.95, repetition penalty=1.1). Each successive generation (Generations 1-4) was initialized from pretrained Qwen3-8B weights and exclusively fine-tuned on synthetic records produced by the previous generation, with no exposure to original human-authored data. This design isolated training data degradation effects from accumulated model drift, enabling direct comparison with GPT-2 experiments while testing whether larger model capacity (8B vs 124M parameters) or parameter-efficient fine-tuning could mitigate degradation dynamics.

### 4.2.2 Unconditional text generation

**GPT-2**

The GPT-2 base architecture was implemented using the Hugging Face Transformers library. The tokenizer was augmented with custom special tokens to demarcate report boundaries, with a maximum sequence length of 512 tokens. Training hyperparameters were consistent across all generations: AdamW optimizer with learning rate $2\times10^{-5}$, batch size 16, 10 epochs, and 100 warmup steps with linear learning rate scheduling. Data were split into 80% training, 10% validation, and 10% test sets using stratified random sampling (random seed 42). Model selection was based on validation loss with early stopping.

Synthetic reports were generated using nucleus sampling[37] with temperature 0.7, top-k 50, and top-p 0.95, with a maximum generation length of 256 tokens. Generation was unconditional, providing only the prompt "<|startoftext|>Findings:" and allowing the model to complete both sections.

Model collapse was quantified across three dimensions. Linguistic diversity metrics included Type-Token Ratio, vocabulary size (excluding stopwords), and n-gram repetition rates (unigrams, bigrams, trigrams) computed using NLTK tokenization. Clinical content degradation was assessed through anatomical and pathological term frequencies using regular expression pattern matching, Findings-Impression alignment using TF-IDF vectorization with cosine similarity, and a three-tier term specificity hierarchy (general, intermediate, specific diagnoses). Condition co-occurrence was analyzed using 10×10 matrices tracking ten pathological conditions (pneumonia, effusion, edema, atelectasis, pneumothorax, consolidation, mass, nodule, fracture, cardiomegaly), with conditional probabilities computed by normalizing co-occurrence counts. Template emergence was quantified by analyzing sentence beginning patterns (first three words) and part-of-speech tag sequences (first five tags per sentence). Readability was assessed using the Flesch Reading Ease score.

**Qwen3-8B**

To validate whether degradation dynamics extend to specialized clinical documentation, we conducted a controlled self-referential training experiment using ophthalmology records. Glaucoma documentation requires precise anatomical terminology and standardized diagnostic codes, making it highly sensitive to loss of clinical specificity. Records comprised 12 standardized sections spanning the complete clinical workflow: Chief Complaint, History of Present Illness, Past Medical History, Family History, Physical Examination, Ancillary Examinations, Laboratory Results, Imaging/Ophthalmic Findings, Surgery/Procedures, Diagnosis, Prescription/Medication, and Treatment/Plan. The dataset was partitioned into training (n=800), validation (n=100), and test (n=100) sets (seed=42), with mean record length of 201±35 words.

We fine-tuned Qwen3-8B using Low-Rank Adaptation (LoRA; r=16, α=32, dropout=0.05) targeting all attention and feed-forward projections. Training employed AdamW optimization (learning rate $3\times10^{-5}$, cosine schedule with 10% warmup, weight decay 0.01) for 5 epochs with early stopping (patience=3) monitoring validation loss. Effective batch size was 16 with bfloat16 precision and gradient checkpointing.

Generation 0 (G0) was fine-tuned on authentic records. Each subsequent generation $G_n$ (n=1,...,4) generated 1,000 synthetic records using $G_{n-1}$ with nucleus sampling (temperature=0.8, top_p=0.95, repetition penalty=1.1), then fine-tuned fresh base model weights exclusively on synthetic data—isolating training data degradation from accumulated model drift.

Evaluation metrics included vocabulary size, type-token ratio, n-gram diversity, domain-specific term frequencies (anatomical structures, diagnostic codes, procedures, medications) normalized per 1,000 words using a curated ophthalmology lexicon, and section completeness rates.

### 4.2.3 Conditional text generation

To evaluate whether providing rich patient-specific context can mitigate or delay degradation induced by self-referential training, we investigated conditional generation of discharge instructions—a safety-critical, patient-facing clinical documentation task. Discharge summaries were obtained from MIMIC-IV-NOTE v2.2, a publicly available database of de-identified clinical notes from Beth Israel Deaconess Medical Center. An initial random sample of 30,000 discharge summaries was extracted for processing. Each summary was parsed to extract structured sections using regex-based pattern matching for standard clinical headers.

The conditional generation task used three input sections as context (Medications on Admission, Discharge Diagnosis, and Discharge Medications) to generate a fourth section (Discharge Instructions). This configuration reflects a clinically meaningful task: generating patient-facing instructions based on the medical context of hospitalization.

Strict token limits of 512 tokens were applied to both context and target sections to ensure consistent sequence lengths across all training generations. After filtering for complete section availability and token length constraints, 11,830 samples remained (44.7% retention), which were split into training (n=9,464; 80%), validation (n=1,183; 10%), and test (n=1,183; 10%) sets using stratified random sampling with a fixed seed (42) for reproducibility.

GPT-2 medium (355 million parameters) served as the base architecture, with the vocabulary extended by seven special tokens to delineate input structure: `<|startoftext|>`, `<|endoftext|>`, `<|context|>`, `<|endcontext|>`, `<|report|>`, `<|endreport|>`, and `<|pad|>`. Token embeddings for added tokens were randomly initialized. The maximum sequence length was set to 1,024 tokens to accommodate combined context and target sequences.

Input sequences were formatted as:

<|startoftext|><|context|>[context-text]<|endcontext|><|report|>[target-text]<|endreport|><|endoftext|>

During training, the loss was computed only on tokens following the `<|report|>` marker, ensuring the model learned to generate clinical content conditioned on the provided context while not being penalized for context reproduction.

Models were trained for 20 epochs per generation using the AdamW optimizer with a learning rate of 5 x 10^-5, weight decay of 0.01, and gradient clipping at norm 1.0. A cosine learning rate schedule with 10% warmup steps was applied. Training used a batch size of 4 per GPU with gradient accumulation over 2 steps (effective batch size of 8), and mixed-precision (FP16) training for computational efficiency. Multi-GPU training was performed using PyTorch DistributedDataParallel. Model selection was based on the lowest validation loss, with checkpointing at each epoch.

Five training generations (Generation 0-4) were conducted to study model collapse. Generation 0 was trained on real clinical data. For subsequent generations (1-4), the model from the previous generation was used to generate synthetic discharge instructions for all training contexts. Synthetic data generation used nucleus sampling with temperature 0.8, top-p 0.9, top-k 50, and repetition penalty 1.1. Generation terminated at the `<|endreport|>` token or a maximum of 512 new tokens. The same validation and test sets (containing real data) were retained across all generations to enable direct comparison.

**Linguistic quality** was assessed using perplexity on the held-out test set (computed as the exponential of cross-entropy loss on target tokens), lexical diversity (unique vocabulary size across all outputs), and n-gram repetition rates (proportion of n-grams appearing more than once, for n=1,2,3).

**Clinical content quality** was evaluated through pattern-based detection of clinical instructions (e.g., "take medication," "follow up," "avoid") versus generic template phrases (e.g., "it was a pleasure," "you were admitted"), reported as rates per 1,000 words. The clinical-to-template ratio quantified the balance between actionable clinical content and boilerplate language.

**Semantic coherence** was measured using 21 metrics including topic similarity (cosine similarity between context and output topic vectors), important term overlap (Jaccard similarity of high-TF-IDF terms), and output grounding (proportion of output medical terms present in context).

## 4.3 Study Design and data sources for medical image to text generation

### 4.3.1 Dataset and Multi-Round Training Framework for visual language model

**To investigate whether visual conditioning could prevent the linguistic and medical knowledge degradation observed in text-only models, we examined multimodal radiology report generation using the MIMIC-CXR dataset.** We utilized a subset of 9,781 chest X-ray images paired with their corresponding clinical reports from the full MIMIC-CXR database, representing diverse pathological presentations across critical care settings. Each image-report pair included structured CheXpert[38] labels quantifying the presence of 14 critical radiographic findings including pneumothorax, pleural effusion, consolidation, edema, and cardiomegaly—conditions requiring immediate clinical attention.

We employed R2GenGPT, a state-of-the-art vision-language model specifically designed for radiology report generation, which integrates a Swin-Transformer-based[39] vision encoder with Llama-2[40] (7B parameters) through an efficient visual alignment module while keeping the language model parameters frozen. Following the same self-referential training protocol established for text-only experiments, we trained five generations: Generation 0 on authentic radiologist-authored reports paired with chest X-rays, with each subsequent generation (1-4) trained exclusively on synthetic reports generated by its predecessor while viewing the same real chest X-ray images. This design isolated the effects of report text collapse while maintaining consistent visual input, enabling assessment of whether grounding in authentic medical images could anchor models to produce clinically accurate text despite training on progressively degraded synthetic reports.

### 4.3.2 Model Architecture

The R2GenGPT architecture was employed for radiology report generation. The model integrates a Swin Transformer (swin-base-patch4-window7-224) as the visual encoder with LLaMA-2 7B as the language model backbone. Visual features extracted from chest radiographs are projected to the language model embedding space through a learned linear projection layer followed by layer normalization. The prompt template "Generate a comprehensive and detailed diagnosis report for this chest X-ray image" guided report generation.

The delta alignment training strategy was adopted, which applies Low-Rank Adaptation (LoRA) exclusively to the visual encoder while keeping the language model frozen. LoRA parameters were set to rank r=16 and scaling factor alpha=16, with dropout of 0.1, targeting the query and value projection matrices. This configuration results in approximately 5 million trainable parameters, enabling efficient fine-tuning while preserving the language model's medical knowledge.

### 4.3.3 Self-referential training Procedure

Model collapse was induced through iterative training on synthetic data across five generations (G0-G4). Generation 0 (G0) was trained on authentic MIMIC-CXR reports. For subsequent generations, the trained model generated synthetic reports for all training images, which then served as the training data for the next generation. This self-referential training procedure was repeated through Generation 4.

Training hyperparameters were held constant across all generations: batch size of 8, validation batch size of 16, learning rate of 1e-4 with cosine annealing, and 20 epochs per generation. Mixed precision training (bfloat16) was employed on 3 NVIDIA RTX 4090 GPUs using the PyTorch Lightning framework with distributed data parallel strategy. Decoding parameters included beam size of 3, repetition penalty of 2.0, length penalty of 2.0, minimum generation length of 80 tokens, and maximum generation length of 120 tokens. Checkpoints were selected based on a weighted combination of BLEU-4 (50%) and CIDEr (50%) scores on the validation set.

### 4.3.4 Evaluation Metrics

**Linguistic Quality:** Report quality was assessed using standard natural language generation metrics: BLEU-1 through BLEU-4 for n-gram precision, ROUGE-L for longest common subsequence overlap, METEOR for semantic similarity incorporating synonymy, and CIDEr for consensus-based evaluation. Lexical diversity was quantified through unique report ratio (proportion of non-duplicate reports), vocabulary size, and trigram entropy.

**Clinical Safety Assessment:** A composite clinical safety score was developed to quantify diagnostic reliability degradation across model generations. The score comprises three independent, equally weighted components (33.3% each): critical finding sensitivity, hallucination rate, and report utility.

Critical finding sensitivity was defined as the mean detection rate across five life-threatening radiological findings: pneumothorax, pleural effusion, consolidation, pulmonary edema, and cardiomegaly. Detection was performed using keyword matching with negation detection (30-character preceding context window), with ground truth labels obtained from CheXpert annotations.

Hallucination rate measured the proportion of reports containing fabricated content, including anatomically impossible descriptions and phantom findings with no corresponding image evidence.

Report utility was a composite of four equally weighted sub-components: linguistic uniqueness (ratio of unique to total reports), artifact-free rate (absence of training artifacts such as loss values or repeated phrases), diagnostic consistency (Cohen's κ between generated and reference pathology mentions), and terminology precision (absence of non-actionable language).

The composite score was calculated as:

Safety Score = 0.33 × Sensitivity + 0.33 × (1 - Hallucination Rate) + 0.33 × Utility

Sensitivity analysis confirmed that the pattern of progressive degradation was robust across alternative weighting schemes, including detection-prioritized (50/30/20) and safety-balanced (45/45/10) configurations (Supplementary). All component scores are bounded [0, 1], with higher values indicating better performance.

## 4.4 Study Design and data sources for medical image generation

### 4.4.1 Dataset and Multi-Round Training Framework

To examine whether self-referential training induces distributional degradation and demographic bias in medical imaging, we utilized 5,534 chest radiographs from the MIMIC-CXR database, standardized to 128×128 pixel grayscale format. We employed an unconditional denoising diffusion probabilistic model[41] (DDPM) with U-Net architecture[42], following the same self-referential training protocol: Generation 0 trained on authentic radiographs, with each subsequent generation trained exclusively on synthetic outputs from its predecessor through Generation 4.

We tracked 11 pathological labels using a pre-trained ResNet classifier, with Cardiomegaly (64.8%), Effusion (58.0%), and Atelectasis (54.3%) most prevalent at baseline. For demographic analysis, Generation 0 exhibited 53.2% male representation with mean age 64.6±17.3 years (range: 18-100), reflecting typical critical care demographics.

### 4.4.2 Diffusion Model Architecture

Unconditional image generation was performed using a Denoising Diffusion Probabilistic Model (DDPM)[41] implemented with a U-Net backbone architecture. The U-Net comprised a base dimension of 64 channels with progressive channel multipliers of (1, 2, 4, 8), resulting in feature maps of 64, 128, 256, and 512 channels at successive resolution levels. The architecture employed residual blocks with RMS normalization, sinusoidal positional embeddings for diffusion timestep conditioning, and linear attention mechanisms. The diffusion process utilised 1,000 forward diffusion timesteps during training with a linear noise schedule, and 250 denoising steps during sampling via DDIM (Denoising Diffusion Implicit Models) acceleration[43].

### 4.4.3 Self-referential Training Procedure

In this paradigm, each generation's model was trained exclusively on synthetic images produced by the immediately preceding generation. Generation 0 (baseline) was trained on the 5,534 real chest radiographs. Generations 1 through 4 were each trained on 5,534 synthetic images sampled from the previous generation's model, creating a chain of purely synthetic inheritance.

All models were trained with identical hyperparameters: batch size of 16, learning rate of $1 \times 10^{-4}$ with Adam optimizer, 100,000 training steps, exponential moving average (EMA) with decay of 0.995, and automatic mixed-precision (AMP) training. Training was performed on NVIDIA GPUs using the Accelerate library for distributed training support.

### 4.4.4 Evaluation Metrics

**Fréchet Inception Distance (FID)[44].** Image quality and distributional fidelity were quantified using FID scores computed between each generation's synthetic output and the original real image distribution (Generation 0). Features were extracted from the average pooling layer of an ImageNet-pretrained Inception-v3 network[45] (2,048-dimensional vectors). To ensure statistical robustness and equal sample sizes across conditions, FID scores were calculated using bootstrap resampling with n = 5,534 images per condition and 10 iterations, reporting mean ± standard deviation.

**Pathology distribution analysis.** Disease label predictions were obtained using a DenseNet-121 classifier pretrained on MIMIC-CXR (torchxrayvision[46], model: densenet121-res224-mimic_ch). This model outputs probability scores for 18 thoracic pathologies. Analysis focused on the 11 clinically relevant labels with non-constant variance across generations: Atelectasis, Cardiomegaly, Consolidation, Edema, Effusion, Enlarged Cardiomediastinum, Fracture, Lung Lesion, Lung Opacity, Pneumonia, and Pneumothorax. Mean prediction probabilities and sample counts (probability threshold > 0.5) were computed for each generation.

### 4.4.5 Physician Evaluation of AI-Generated Discharge Summaries

To assess the real-world clinical utility of AI-generated discharge summaries under self-referential training, we conducted a physician-in-the-loop evaluation simulating routine discharge documentation workflows. Two independent physician annotators (board-certified clinicians) evaluated a stratified sample of 100 AI-generated discharge summaries comprising 25 unique patient cases, each with outputs from four generations (Gen1 through Gen4). Generation 1 was trained on authentic human-authored discharge summaries from MIMIC-IV; generations 2 through 4 were trained recursively on synthetic outputs from the preceding generation, following the iterative synthesis procedure described previously.

For each AI-generated summary, annotators manually edited the text to achieve clinical acceptability—correcting medical inaccuracies, restoring missing critical information, improving logical flow, and ensuring patient safety standards. This editing paradigm mirrors real-world clinical workflow where physicians review AI-generated documentation before patient discharge. Annotators were blinded to generation labels to prevent bias.

Four complementary metrics quantified clinical quality degradation:

- **Edit distance percentage:** The proportion of characters changed during physician editing (Levenshtein distance normalized by original text length), with higher values indicating greater deviation from clinically acceptable content.
- **AI content retention:** The percentage of original AI-generated content preserved after physician editing, serving as a direct measure of clinical utility.
- **Word error rate:** The density of word-level corrections per generated word, providing granular assessment of linguistic accuracy.
- **Editing time:** The duration (seconds) physicians spent reviewing and correcting each summary, reflecting cognitive burden imposed by AI errors.

## 4.5 Mitigation strategies for model collapse

We systematically evaluated three mitigation strategies to counteract model collapse in clinical text generation. All experiments used GPT-2 (124M parameters) fine-tuned on 5,000 radiology reports sampled from MIMIC-CXR, with training conducted over four successive generations (Gen-0 through Gen-4). To ensure valid comparisons, all conditions shared a common Gen-0 model trained on identical real data, with hyperparameters held constant (learning rate $2\times10^{-5}$, batch size 16, 10 epochs, maximum sequence length 512 tokens). Generation used nucleus sampling (temperature 0.7, top-k 50, top-p 0.95).

### 4.5.1 Mixed training with real data

We hypothesized that incorporating real data during self-referential training could preserve linguistic diversity by continuously injecting authentic patterns otherwise lost through synthetic-only training. We tested four mixing ratios: control (0% real, pure synthetic), mixed_25 (25% real), mixed_50 (50% real), and mixed_75 (75% real). At each generation, real samples were freshly drawn from the base dataset while synthetic samples came from the previous generation's model output. Total training set size remained fixed at 5,000 reports regardless of mixing ratio, isolating the effect of data composition.

### 4.5.2 Increasing synthetic data volume provides minimal protection

We tested whether model collapse results from insufficient training signals that could be compensated by increasing synthetic data volume. This hypothesis posits that larger datasets might preserve distributional diversity by providing more examples of rare cases, even when all data are synthetic.

**Clinical text generation.** The control condition maintained fixed volume (5,000 samples per generation), while the increasing-volume condition scaled linearly: 10,000 samples at Generation 1, 15,000 at Generation 2, 20,000 at Generation 3, and 25,000 at Generation 4, representing a 2× to 5× increase over baseline. Both conditions used pure synthetic data (0% real data mixing), enabling direct comparison between quantity-based and composition-based mitigation approaches.

**Medical image synthesis.** We applied an equivalent scaling protocol to chest X-ray generation. The control condition maintained fixed volume (500 images per generation), while the increasing-volume condition scaled linearly: 1,000 images at Generation 1, 1,500 at Generation 2, 2,000 at Generation 3, and 2,500 at Generation 4. This matched the 2× to 5× scaling factor used in text experiments, enabling cross-modal comparison of volume-based mitigation efficacy.

All other training hyperparameters remained identical to the pure synthetic baseline conditions described in Sections 4.2 and 4.4.

### 4.5.3 Quality-aware filtering enhances real data efficiency

We developed modality-specific filtering strategies to optimize the selection of both synthetic and real data samples, motivated by the observation that model collapse preferentially eliminates rare findings while synthetic outputs converge toward high-frequency prototypical patterns. The underlying principle was to retain synthetic samples most resembling authentic data (preserving generation quality) while selecting real samples capturing distributional tails (preserving pathological variability).

**Image quality filtering.** For medical image synthesis, we quantified sample quality using Mahalanobis distances computed within a perceptual-quality embedding space. This embedding combined three feature types: (1) 2,048-dimensional Inception-v3 features used for FID computation, (2) pixel-intensity variance capturing radiographic contrast, and (3) Sobel edge-content descriptors quantifying anatomical boundary preservation. Synthetic samples falling in the lowest 25th percentile of this composite distribution—representing redundant, low-information, or degraded images—were excluded prior to real-synthetic mixing.

**Text quality filtering.** For clinical text generation, we implemented an embedding-based filtering approach using GPT-2 Large (774M parameters) as an external reference model, thereby avoiding the circular bias that would arise from using the Generation-0 model to score its own descendants. We extracted 1,280-dimensional embeddings via mean pooling over non-padding token positions and computed k-nearest neighbor distances (k=10, cosine similarity metric) to the authentic data distribution. Synthetic samples with low k-NN distances (closest to the real distribution, indicating high fidelity) were preferentially retained, while real samples with high k-NN distances (most isolated from distribution modes, representing diverse edge cases) were preferentially selected. This complementary selection strategy aimed to preserve generation quality while maintaining the pathological variability essential for safe medical AI.

Experimental controls. All filtering conditions maintained a fixed 25% real / 75% synthetic mixing ratio, enabling controlled comparison with the unfiltered mixed training strategy (Section 4.5.1). The synthetic candidate pool remained fixed at 5,000 samples across all strategies, isolating the effect of selection method from pool size. An alternative perplexity-based filtering approach is described in Supplementary Methods.

## 4.6 Statistical analysis

**Descriptive Statistics.** Continuous variables are presented as means with standard deviations (mean ± SD) or 95% confidence intervals (CI) as specified. For normally distributed data, 95% CIs were calculated as mean ± 1.96 × (SD/$\sqrt{n}$). For metrics with non-normal distributions or small samples, 95% CIs were calculated via bootstrap resampling with 10,000 iterations unless otherwise specified.

**Image Quality Assessment.** Fréchet Inception Distance (FID) scores were computed using bootstrap resampling (n = 1,384 images per condition, 10 iterations) to ensure equal sample sizes across conditions while providing uncertainty estimates, reported as mean ± SD.

**Demographic Bias Quantification.** Age distribution shifts were quantified using the Wasserstein distance (earth mover's distance) between generation-specific distributions and the Generation 0 baseline. Gender distribution changes were assessed using chi-square tests comparing observed frequencies against baseline proportions, with statistical significance defined as $P < 0.05$.

**Physician Evaluation Study.** The physician evaluation followed a repeated-measures design in which the same 25 patient cases were evaluated across all four generations by two independent annotators, yielding 100 paired observations (25 patients × 4 generations). Inter-rater reliability was assessed using the Intraclass Correlation Coefficient (ICC) with a two-way random effects model for absolute agreement, denoted ICC(2,1). ICC values were interpreted following standard thresholds: ≥0.75 excellent, 0.60–0.74 good, 0.40–0.59 fair, and <0.40 poor. Pearson correlation coefficients were computed to assess linear association between annotator scores. Consensus scores were calculated as the mean of both annotators' assessments. Generation-level summary statistics are reported with 95% CIs calculated using the t-distribution with $n-1$ degrees of freedom (n = 25 patients per generation).

**Diagnostic Agreement.** Cohen's kappa (κ) was used to quantify diagnostic consistency between model generations for vision–language experiments, with interpretation following standard thresholds: <0.20 slight, 0.21–0.40 fair, 0.41–0.60 moderate, 0.61–0.80 substantial, and 0.81–1.00 almost perfect agreement.

**Correlation Analysis.** Pearson correlation coefficients were used for normally distributed continuous variables; Spearman rank correlation (ρ) was used for non-parametric associations or when assessing monotonic relationships.

Statistical significance was defined as two-tailed $P < 0.05$ for all analyses. All analyses were performed using Python v.3.9 with SciPy v.1.7.3, statsmodels, and scikit-learn libraries.

# Code and data availability

Analysis code and synthetic datasets are available upon publication. Original medical datasets require appropriate data use agreements from respective sources. We provide detailed reproduction instructions and pre-trained models for research purposes.